\documentclass[prb,twocolumn,superscriptaddress]{revtex4}

\usepackage{pslatex}
\usepackage{amsmath}
\usepackage{amssymb}
\usepackage{graphicx}
\usepackage{pstricks}
\usepackage{wasysym}

\begin{document}

\bibliographystyle{apsrev}

\title{Non-Abelian topological phases in an extended Hubbard model}



\author{Michael Freedman}
\affiliation{Microsoft Research, One Microsoft Way,
Redmond, WA 98052}

\author{Chetan Nayak}
\affiliation{Microsoft Research, One Microsoft Way,
Redmond, WA 98052}
\affiliation{Department of Physics and
  Astronomy, University of California, Los Angeles, CA 90095-1547}
\author{Kirill Shtengel}
\affiliation{Microsoft Research, One Microsoft Way,
Redmond, WA 98052}
\date{\today}

\begin{abstract}
  We describe four closely related Hubbard-like models (models
  $A$,$B$,$C$ and $D$) of particles that can hop on a $2D$ Kagom\'{e}
  lattice interacting via Coulomb repulsion. The particles can be
  either bosons (models $A$ and $B$) or (spinless) fermions (models
  $C$ and $D$). Models $A$ and $C$ also include a ring exchange term.
  In all four cases we solve equations in the model parameters to
  arrive at an exactly soluble point whose ground state manifold is
  the extensively degenerate ``$d$-isotopy space'' $\overline{V}_d$,
  $0<d<2$. Near the ``special'' values, $d = 2 \cos \pi/k+2$,
  $\overline{V}_d$ should collapse to a stable topological phase with
  anyonic excitations closely related to $\text{SU}(2)$ Chern-Simons
  theory at level $k$. We mention simplified models $A^-$ and $C^-$
  which may also lead to these topological phases.
\end{abstract}

\pacs{}
\maketitle

\section{Introduction}

Since the discovery of the fractional quantum Hall effect in 1982
\cite{Tsui82}, topological phases of electrons have been a subject of
great interest.  Many abelian topological phases have been discovered
in the context of the quantum Hall regime \cite{DasSarma97}. More
recently, high-temperature superconductivity
\cite{Anderson87,Kivelson87,Kalmeyer87,Laughlin88a,Laughlin88b,Fetter89,Chen89,Read89c,Read91a,Read91b,Wen91b,Mudry94a,Balents98,Senthil00}
and other complex materials have provided the impetus for further
theoretical studies of and experimental searches for abelian
topological phases.
The types of microscopic models admitting such phases is
now better understood \cite{Moessner01,Balents02,Senthil02,Motrunich02}.

Much less is known about non-abelian topological phases. They are
reputed to be obscure and complicated, and there has been little
experimental motivation to consider non-abelian topological phases,
apart from some tantalizing hints that the quantum Hall plateau
observed at $\nu=5/2$ might be non-abelian
\cite{Willett87,Pan99,Moore91,Greiter92,Nayak96c,Read96,Fradkin98}.
However, non-abelian topological states would be an attractive milieu
for quantum computation \cite{Kitaev97,Freedman01}.  Furthermore, a better
understanding of non-abelian topological phases would help determine
if they have already been unwittingly observed in nature.  It is,
therefore, encouraging that recent progress in understanding the
Hamiltonian formulation of a class of topological field theories has
reduced computations in these theories to combinatorial-pictorial
manipulations \cite{Kauffman94,Freedman03}.  These theories
do not break parity or time-reversal symmetry, which may
have practical advantages. A further virtue of this
formulation of these theories is that it exposes a strategy for
constructing microscopic physical models which admit the corresponding
phases; since Hilbert space is reduced to a set of pictorial rules,
the models should impose these rules as energetically-favorable
conditions satisfied by the ground state. In this paper, we show how
this approach can be implemented.

The microscopic models which we construct are `quasi-realistic'.
Their precise form is not likely to be realized in any material.
However, they are soluble, so we know that they do, indeed, support
the non-abelian topological states of matter which interest us, and
one can imagine a real material whose Hamiltonian can be viewed as a
small perturbation of one of the Hamiltonians of this paper.  It may
also be possible to design a quantum dot or Josephson junction array
with such a Hamiltonian.

The non-abelian topological phases which arise in this paper are
related to the doubled $\text{SU}(2)_k$ Chern-Simons theories
described in Refs.~\onlinecite{Freedman03,Good_guys}.  In
appendix~\ref{appendix1} some further information about these
theories, called $\mathcal{D}K_k$, is given \footnote{This
  terminology, first used in Ref.~ \onlinecite{Freedman03}, indicates
  the Drinfeld double of the Kauffman bracket version of the
  $\text{SU}(2)$ topological quantum field theory (TQFT) at level $k$.
  The Kauffman bracket version neglects the $-1$'s derived from the
  Frobenius-Schur indicator.}.  For the moment, we note that
$\mathcal{D}K_k$ has ground state degeneracy $(k+1)^2$ on the torus
$T^2$ and should be viewed as a natural family containing the
topological (deconfined) phase of $\text{Z}_2$ gauge theory as its
initial element, $k=1$. For $k\geq 2$ the excitations are non-abelian.
For $k=3$ and $k\geq 5$ the excitations are computationally universal
\cite{FLW02a,FLW02b}. In Appendix~\ref{appendix2} some general purpose
formulae for the perturbation theory of Hubbard models are derived.

The basic strategy of the present paper is to construct a Hamiltonian
which enforces a subset -- $d$-isotopy \cite{Freedman03,Good_guys} --
of the defining conditions of the Hilbert space of these theories. The
resulting low-energy subspace will be called $\overline{V}_d$.  It has
been argued \cite{Freedman03,Good_guys} that for special values of the
parameter $d$, namely $d=2\cos(\pi/k+2)$, $\overline{V}_d$ will
collapse to the stable topological phase $\mathcal{D}K_k$.

\section{The basic model}
\label{sec:model}

\begin{figure}[hbt]
\includegraphics[width=2.5in]{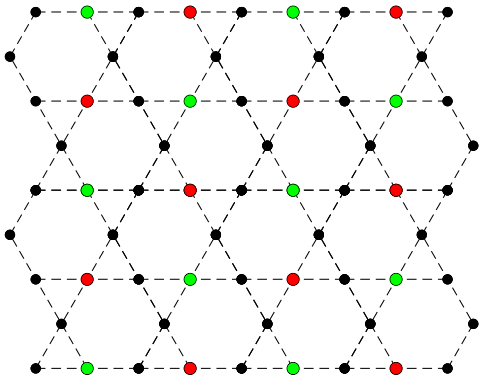}
\caption{Kagom\'{e} lattice $\mathcal{K}$
  with the special sublattices $\mathcal{R}$ (depicted red) and
  $\mathcal{G}$ (green).}
\label{fig:Kagome}
\end{figure}

The models described in this paper, both bosonic and fermionic, are
defined on the Kagom\'{e} lattice. The basic lattice is shown in
Fig.~\ref{fig:Kagome}. The sites of the lattice are not completely
equivalent, in particular we choose two special sublattices -
$\mathcal{R}$ (red) and $\mathcal{G}$ (green) whose significance shall
became clear later on.  In some cases we will introduce ``defects'' in
this sublattice arrangement.

Our generic Hamiltonian is given by:
\begin{multline}
  \label{eq:Hubbard}
  H =  \sum_i \mu_i n_i +  U_0 \sum_i n^2_i
  +  U \sum_{(i,j)\in \hexagon} n_i n_j
  + \sum_{(i,j)\in \bowtie, \notin \hexagon } V_{i j} n_i n_j\\
   - \sum_{\langle i,j \rangle} t_{ij}
  (c^\dag_i c_j + c^\dag_j c_i ) + \text{Ring}.
\end{multline}
Here $n_i \equiv c_i^\dag c_i$ is the occupation number on site $i$,
$\mu_i$ is the corresponding chemical potential. $U_0$ is the usual
onsite Hubbard energy $U_0$ (clearly superfluous for spinless
fermions). $U$ is a (positive) Coulomb penalty for having two
particles on the same hexagon while $V_{ij}$ represent a penalty for
two particles occupying the opposite corners of ``bow ties'' (in other
words, being next-nearest neighbours on one of the straight lines). We
allow for the possibility of inhomogeneity so not all $V_{ij}$ are
assumed equal.  Specifically, define $v^{c}_{ab}= V_{ij}$ where $a$ is
the color of site $(i)$, $b$ is the colour of $(j)$, and $c$ is the
colour of the site between them. In the lattice in
Fig.~\ref{fig:Kagome} we have, possibly distinct, $v^{g}_{bb}$,
$v^{b}_{bb}$, $v^{g}_{bg}$, $v^{b}_{rb}$, and $v^{b}_{rg}$, where $r
\in \mathcal{R}$, $g \in \mathcal{G}$, and $b \in \mathcal{B}=
\mathcal{K} \diagdown (\mathcal{R} \cup \mathcal{G})$.  $t_{ij}$ is
the usual nearest-neighbour tunnelling amplitude which is also assumed
to depend only on the colour of the environment: $t_{ij}\equiv
t^{c}_{ab}$ where $c$ now refers to the colour of the third site in a
triangle.  Finally, ``Ring'' is a ring exchange term -- an additional
kinetic energy term which we add to the Hamiltonian on an \emph{ad
  hoc} basis to allow correlated multi-particle hops which ``shift''
particles along some closed paths (see more on this term below).  Ring
exchange terms can be justified semiclassically, but are not generally
included as bare terms in the Hubbard model \footnote{Of course, small
  ring terms (or at least their kinetic parts) arise perturbatively --
  see Appendix~\ref{appendix2}. E.g. a four-particle move described
  later in the paper as a ``bow tie'' move occurs at order 4.}. In
models $B$ and $D$ we remove the Ring term at the expense of
complicating the sublattice $\mathcal{R}$ (and thus the interactions)
with the addition of several new colors.

The onsite Hubbard energy $U_0$ is considered to be the biggest energy
in the problem, and we shall set it to infinity, thereby restricting
our attention to the low-energy manifold with sites either unoccupied
or singly-occupied. The rest of the energies satisfy the following
relations: $U\gg t_{ij},V_{ij}, \mu_{i}$; we shall be more specific
about relations between various $t_{ij}$'s, $V_{ij}$'s and
$\mu_i$'s later.

\begin{figure}[bht]
\includegraphics[width=2.5in]{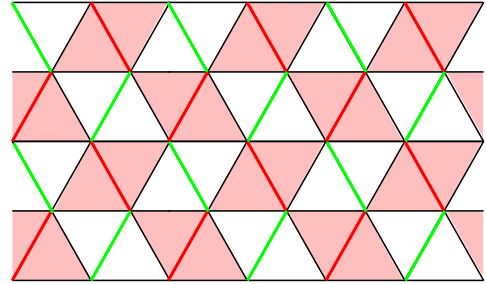}
\caption{
  Triangular lattice $\mathcal{T}$ obtained from $\mathcal{K}$ by
  connecting the centers of adjacent hexagons. The bonds corresponding
  to the special sublattices $\mathcal{R}$ and $\mathcal{G}$ are shown
  in red and green correspondingly. See text for more on bond
  color-coding. Triangles with one red side are shaded as guide to
  the eyes: these triangles will be essential for the dynamics of our
  models.}
\label{fig:Triang}
\end{figure}

We are particularly interested in the 1/6-filled case (i.e.
$N_p\equiv\sum_i n_i = N/6$, where N is the number of sites in the
lattice).
The lowest-energy band then consists of configurations in which there
is exactly one particle per hexagon, hence all $U$-terms are set to
zero. (One example of such a configuration is given by particles
occupying all sites of sublattice $\mathcal{R}$.) These states are
easier to visualise if we consider a triangular lattice $\mathcal{T}$
whose sites coincide with the centers of hexagons of $\mathcal{K}$.
($\mathcal{K}$ is a \emph{surrounding} lattice for $\mathcal{T}$.)
Then a particle on $\mathcal{K}$ is represented by a dimer on
$\mathcal{T}$ connecting the centers of two adjacent hexagons of
$\mathcal{K}$.  The condition of one particle per hexagon translates
into the requirement that no dimers share a site.  In the 1/6-filled
case this low-energy manifold coincides with the set of all dimer
coverings (perfect matchings) of $\mathcal{T}$.  The ``red'' bonds of
$\mathcal{T}$ (the ones corresponding to the sites of sublattice
$\mathcal{R}$) themselves form one such dimer covering, a so-called
``staggered configuration''. This particular covering is special: it
contains no ``flippable plaquettes'', or rhombi with two opposing
sides occupied by dimers (see Fig.~\ref{fig:Triang}).

So henceforth particles live on \emph{bonds} of the triangular lattice
(Fig.~\ref{fig:Triang}) and are represented as dimers \footnote{The
  important feature of the triangular lattice, for us, will be that if
  is \emph{not} bipartite. On the edges of a bipartite lattice, our
  models will have an additional, undesired, conserved quantity
  (integral winding numbers), so the triangular lattice gives the
  simplest realization.}. In particular, a particle hop corresponds to
a dimer ``pivoting'' by $60^\circ$ around one of its endpoints, $
V_{ij}=v^{c}_{ab}$ is now a potential energy of two parallel dimers on
two opposite sides of a rhombus (with $c$ being the color of its short
diagonal).

From the relation to dimer coverings exhibited above, it is clear that
our model is in the same family as the quantum dimer model
\cite{Rokhsar88}, which has recently been shown to have an abelian
topological phase on the triangular lattice \cite{Moessner01} which,
in the notation of this paper, is $\mathcal{D}K_1$.  Here, we show how
other values of $k$ can be obtained.

The goal now is to derive the effective Hamiltonian acting on this
low-energy manifold represented by all possible dimer coverings of
$\mathcal{T}$.  Our analysis is perturbative in ${t}/{U}=: \epsilon$.

The initial, unperturbed ground state manifold for $U_0=\infty$, $U$
large and positive, all $t_{ij},\; V_{ij}=0$ and all $\mu_i$
equal is spanned by the dimerizations $\mathcal{D}$ of the triangular
lattice $\mathcal{T}$.  As we gradually turn on the $t$'s, $v$'s, and
Ring terms, we shall see what equations they should satisfy so that
the effective Hamiltonian on $\mathcal{D}$ has the desired space
$\overline{V}_d$ as its ground state manifold (GSM).

Since a single tunnelling event in $\mathcal{D}$ always leads to
dimer ``collisions'' (two dimers sharing an endpoint) with
energy penalty $U$, the lowest order at which the tunnelling
processes contribute to the effective low-energy Hamiltonian is 2.  At
this order, the tunnelling term leads to two-dimer ``plaquette
flips'' as well as renormalisation of bare onsite potentials
$\mu_i$'s due to dimers pivoting out of their positions and back.
We always recompute bare potentials $\mu_i$'s to maintain equality
up to errors $\mathcal{O} (\epsilon^3 )$ among the renormalized
$\widetilde{\mu}_i$'s.  This freedom to engineer the chemical
potential landscape to balance kinetic energy is essential to finding
our ``exactly soluble point'' in the model. Before
we derive the constraints which we encounter in tuning the ground state
manifold (GSM) to $\overline{V}_d$, we must first explain $\overline{V}_d$
and how we map it to dimer coverings.

\section{$d$-isotopy and its local subspaces}
\label{sec:d-isotopy}

\begin{figure}[hbt!]
\includegraphics[width=2.75in]{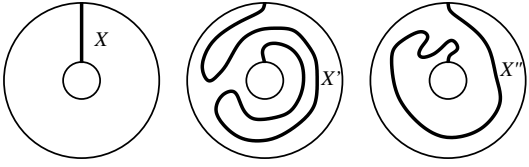}
\caption{Isotopy on
  an annulus: $X\sim X'$ but $X \nsim X''$.}
\label{fig:isotopy}
\end{figure}

Although we will eventually be considering a system on a lattice, it
is useful to begin by defining the Hilbert spaces of interest,
$\overline{V}_d$ and ${V}_d$, in the smooth, lattice-free setting.
Consider a compact surface $Y$ and the set $S$ of all multiloops
\footnote{A multiloop is a collection of non-intersecting loops and
  arcs on a surface. The end points of an arc are required to lie on
  the boundary od the surface, see Ref.~\onlinecite{Good_guys}.}
$X\subset Y$.  If $\partial Y$ (the boundary of $Y$) is non-empty, we
fix once and for all a finite set $P$ of points on $\partial Y$ with
$X \cap \partial Y =P$.  We assume $Y$ is oriented but $X$ should
\emph{not} be. There is a large vector space $\mathbb{C}^S$, of
complex valued functions on $S$.  We say $X$ and $X'$ are
\emph{isotopic} ($X{\sim}X'$) if one may be gradually deformed into
the other with, of course, the deformation being the identity on
$\partial Y$ (see Fig.~\ref{fig:isotopy}).

We may view the isotopy relation as a family of linear constraints on
$\mathbb{C}^S$, namely $\Psi(X)-\Psi(X') =0$ if $X \sim X'$.  The
subspace satisfying these linear constraints is now only of countable
dimension; it consists of those functions which depend only of the
isotopy class $[X]$ of $X$ and may be identified with
$\mathbb{C}^{[S]}$, where $[S]$ is the isotopy classes of multiloops
with fixed boundary conditions.  Note that since all isotopes can be
made by composing small locally-supported ones, so the relations we
just imposed are ``local'' in the sense that we will be able to
implement them with purely local terms in the Hamiltonian.

Let us go further and define an additional local relation which, when
added to isotopy, constitutes the ``$d$-isotopy'' relation.  This
relation is:
\begin{equation}
  \label{eq:d-isotopy}
  d\;\Psi(X)-\Psi(X \cup \bigcirc)=0
\end{equation}

It says that if two multiloops are identical except for the presence
of a small (or, it follows, any contractible) circle, then their
function values differ by a factor of $d$, a fixed positive real
number. In cases of interest to us $1\leq d<2$, so our function is
either neutral to or ``likes'' small circles. We call the subspace
obeying all these constraints the $d$--isotopy space of $Y$ (with
fixed boundary conditions) and write it as $\overline{V}_d \subset
\mathbb{C}^{[S]} \subset\mathbb{C}^S$. The subspace $\overline{V}_d
(T^2)$ is still of countable dimension, or extensively degenerate, on
the torus $T^2$.

It is a remarkable fact (see
Refs.~\onlinecite{Freedman03,FNWW,Good_guys}) that it is very
difficult to add any further local relations to $d-$isotopy without
killing the vector space entirely.  For $\alpha$ a root of unity $\neq
\pm 1$, and $d = \alpha +\overline{\alpha}$ there is such a local
relation, but in almost all cases the natural inner product on
$\overline{V}_d$ fails to be positive definite.  The physically
interesting cases reduce to $\alpha = \text{e}^{\pi \text{i}/(k+2)}$,
$k=1,2,3, \ldots$.  We call the corresponding $d$'s ``special''.

In these cases the local relations are essentially the Jones-Wenzl
idempotents:\\
$k=1$:
\begin{equation}
  \label{eq:JW1}
  \pspicture[0.4](1.0,1.0)
  \psframe[linewidth=0.5pt](0,0)(1.0,1.0)
  \psbezier[linewidth=1.0pt](0.333333333333333,0)
  (0.333333333333333,0.5)(0.333333333333333,0.5)(0.333333333333333,1.0)
  \psbezier[linewidth=1.0pt](0.666666666666667,0)(0.666666666666667,0.5)
  (0.666666666666667,0.5)(0.666666666666667,1.0)
  \endpspicture
  \; - \; \frac{1}{d}
  \pspicture[0.4](1.0,1.0)
  \psframe[linewidth=0.5pt](0,0)(1.0,1.0)
  \psbezier[linewidth=1.0pt](0.333333333333333,0)
  (0.333333333333333,0.333333333333333)
  (0.666666666666667,0.333333333333333)(0.666666666666667,0)
  \psbezier[linewidth=1.0pt](0.666666666666667,1.0)
  (0.666666666666667,0.666666666666667)
  (0.333333333333333,0.666666666666667)(0.333333333333333,1.0)
  \endpspicture
  =0 
\end{equation}
$k=2$:
\begin{multline}
  \label{eq:JW2}
  \pspicture[0.4](1.0,1.0)
  \psframe[linewidth=0.5pt](0,0)(1.0,1.0)
  \psbezier[linewidth=1.0pt](0.25,0)(0.25,0.5)(0.25,0.5)(0.25,1.0)
  \psbezier[linewidth=1.0pt](0.5,0)(0.5,0.5)(0.5,0.5)(0.5,1.0)
  \psbezier[linewidth=1.0pt](0.75,0)(0.75,0.5)(0.75,0.5)(0.75,1.0)
  \endpspicture
  + 
  \frac{1}{d^2-1}
  \left(\: \pspicture[0.4](1.0,1.0)
    \psframe[linewidth=0.5pt](0,0)(1.0,1.0)
    \psbezier[linewidth=1.0pt](0.25,0)(0.25,0.25)(0.5,0.25)(0.5,0)
    \psbezier[linewidth=1.0pt](0.75,0)(0.75,0.5)(0.25,0.5)(0.25,1.0)
    \psbezier[linewidth=1.0pt](0.75,1.0)(0.75,0.75)(0.5,0.75)(0.5,1.0)
    \endpspicture
    + \pspicture[0.4](1.0,1.0)
    \psframe[linewidth=0.5pt](0,0)(1.0,1.0)
    \psbezier[linewidth=1.0pt](0.25,0)(0.25,0.5)(0.75,0.5)(0.75,1.0)
    \psbezier[linewidth=1.0pt](0.5,0)(0.5,0.25)(0.75,0.25)(0.75,0)
    \psbezier[linewidth=1.0pt](0.5,1.0)(0.5,0.75)(0.25,0.75)(0.25,1.0)
    \endpspicture
    \:\right)
  \\
  -  \frac{d}{d^2-1}
  \left(\: \pspicture[0.4](1.0,1.0)
    \psframe[linewidth=0.5pt](0,0)(1.0,1.0)
    \psbezier[linewidth=1.0pt](0.25,0)(0.25,0.25)(0.5,0.25)(0.5,0)
    \psbezier[linewidth=1.0pt](0.75,0)(0.75,0.5)(0.75,0.5)(0.75,1.0)
    \psbezier[linewidth=1.0pt](0.5,1.0)(0.5,0.75)(0.25,0.75)(0.25,1.0)
    \endpspicture
    + \pspicture[0.4](1.0,1.0)
    \psframe[linewidth=0.5pt](0,0)(1.0,1.0)
    \psbezier[linewidth=1.0pt](0.25,0)(0.25,0.5)(0.25,0.5)(0.25,1.0)
    \psbezier[linewidth=1.0pt](0.5,0)(0.5,0.25)(0.75,0.25)(0.75,0)
    \psbezier[linewidth=1.0pt](0.75,1.0)(0.75,0.75)(0.5,0.75)(0.5,1.0)
    \endpspicture
    \:\right)
  = 0
\end{multline}
$k=3$:
\begin{multline}
  \label{eq:JW3}
  \pspicture[0.4](0.9,0.9)
  \scalebox{0.9}{
    \psframe[linewidth=0.5pt](0,0)(1.0,1.0)
    \psbezier[linewidth=1.0pt](0.2,0)(0.2,0.5)(0.2,0.5)(0.2,1.0)
    \psbezier[linewidth=1.0pt](0.4,0)(0.4,0.5)(0.4,0.5)(0.4,1.0)
    \psbezier[linewidth=1.0pt](0.6,0)(0.6,0.5)(0.6,0.5)(0.6,1.0)
    \psbezier[linewidth=1.0pt](0.8,0)(0.8,0.5)(0.8,0.5)(0.8,1.0)
  }
  \endpspicture
  - \frac{d}{d^2-2}
  \pspicture[0.4](0.9,0.9)
  \scalebox{0.9}{
    \psframe[linewidth=0.5pt](0,0)(1.0,1.0)
    \psbezier[linewidth=1.0pt](0.2,0)(0.2,0.5)(0.2,0.5)(0.2,1.0)
    \psbezier[linewidth=1.0pt](0.4,0)(0.4,0.2)(0.6,0.2)(0.6,0)
    \psbezier[linewidth=1.0pt](0.8,0)(0.8,0.5)(0.8,0.5)(0.8,1.0)
    \psbezier[linewidth=1.0pt](0.6,1.0)(0.6,0.8)(0.4,0.8)(0.4,1.0)
  }
  \endpspicture
  - \frac{d^2-1}{d^3-2d}
  \left(\: \pspicture[0.4](0.9,0.9)
    \scalebox{0.9}{
      \psframe[linewidth=0.5pt](0,0)(1.0,1.0)
      \psbezier[linewidth=1.0pt](0.2,0)(0.2,0.2)(0.4,0.2)(0.4,0)
      \psbezier[linewidth=1.0pt](0.6,0)(0.6,0.5)(0.6,0.5)(0.6,1.0)
      \psbezier[linewidth=1.0pt](0.8,0)(0.8,0.5)(0.8,0.5)(0.8,1.0)
      \psbezier[linewidth=1.0pt](0.4,1.0)(0.4,0.8)(0.2,0.8)(0.2,1.0)
    }
    \endpspicture
    + \pspicture[0.4](0.9,0.9)
    \scalebox{0.9}{
      \psframe[linewidth=0.5pt](0,0)(1.0,1.0)
      \psbezier[linewidth=1.0pt](0.2,0)(0.2,0.5)(0.2,0.5)(0.2,1.0)
      \psbezier[linewidth=1.0pt](0.4,0)(0.4,0.5)(0.4,0.5)(0.4,1.0)
      \psbezier[linewidth=1.0pt](0.6,0)(0.6,0.2)(0.8,0.2)(0.8,0)
      \psbezier[linewidth=1.0pt](0.8,1.0)(0.8,0.8)(0.6,0.8)(0.6,1.0)
    }
    \endpspicture
    \:\right)
  \\
  + \frac{1}{d^2-2}
  \left(\: \pspicture[0.4](0.9,0.9)
    \scalebox{0.9}{
      \psframe[linewidth=0.5pt](0,0)(1.0,1.0)
      \psbezier[linewidth=1.0pt](0.2,0)(0.2,0.2)(0.4,0.2)(0.4,0)
      \psbezier[linewidth=1.0pt](0.6,0)(0.6,0.5)(0.2,0.5)(0.2,1.0)
      \psbezier[linewidth=1.0pt](0.8,0)(0.8,0.5)(0.8,0.5)(0.8,1.0)
      \psbezier[linewidth=1.0pt](0.6,1.0)(0.6,0.8)(0.4,0.8)(0.4,1.0)
    }
    \endpspicture
    + \pspicture[0.4](0.9,0.9)
    \scalebox{0.9}{
      \psframe[linewidth=0.5pt](0,0)(1.0,1.0)
      \psbezier[linewidth=1.0pt](0.2,0)(0.2,0.5)(0.6,0.5)(0.6,1.0)
      \psbezier[linewidth=1.0pt](0.4,0)(0.4,0.2)(0.6,0.2)(0.6,0)
      \psbezier[linewidth=1.0pt](0.8,0)(0.8,0.5)(0.8,0.5)(0.8,1.0)
      \psbezier[linewidth=1.0pt](0.4,1.0)(0.4,0.8)(0.2,0.8)(0.2,1.0)
    }
    \endpspicture
    + \pspicture[0.4](0.9,0.9)
    \scalebox{0.9}{
        \psframe[linewidth=0.5pt](0,0)(1.0,1.0)
        \psbezier[linewidth=1.0pt](0.2,0)(0.2,0.5)(0.2,0.5)(0.2,1.0)
        \psbezier[linewidth=1.0pt](0.4,0)(0.4,0.2)(0.6,0.2)(0.6,0)
        \psbezier[linewidth=1.0pt](0.8,0)(0.8,0.5)(0.4,0.5)(0.4,1.0)
        \psbezier[linewidth=1.0pt](0.8,1.0)(0.8,0.8)(0.6,0.8)(0.6,1.0)
      }
      \endpspicture
      + \pspicture[0.4](0.9,0.9)
      \scalebox{0.9}{
        \psframe[linewidth=0.5pt](0,0)(1.0,1.0)
        \psbezier[linewidth=1.0pt](0.2,0)(0.2,0.5)(0.2,0.5)(0.2,1.0)
        \psbezier[linewidth=1.0pt](0.4,0)(0.4,0.5)(0.8,0.5)(0.8,1.0)
        \psbezier[linewidth=1.0pt](0.6,0)(0.6,0.2)(0.8,0.2)(0.8,0)
        \psbezier[linewidth=1.0pt](0.6,1.0)(0.6,0.8)(0.4,0.8)(0.4,1.0)
      }
      \endpspicture
      \:\right) 
    \\  
    - \frac{1}{d^3-2d}
    \left(\: \pspicture[0.4](0.9,0.9)
      \scalebox{0.9}{
        \psframe[linewidth=0.5pt](0,0)(1.0,1.0)
        \psbezier[linewidth=1.0pt](0.2,0)(0.2,0.2)(0.4,0.2)(0.4,0)
        \psbezier[linewidth=1.0pt](0.6,0)(0.6,0.5)(0.2,0.5)(0.2,1.0)
        \psbezier[linewidth=1.0pt](0.8,0)(0.8,0.5)(0.4,0.5)(0.4,1.0)
        \psbezier[linewidth=1.0pt](0.8,1.0)(0.8,0.8)(0.6,0.8)(0.6,1.0)
      }
      \endpspicture
      + \pspicture[0.4](0.9,0.9)
      \scalebox{0.9}{
        \psframe[linewidth=0.5pt](0,0)(1.0,1.0)
        \psbezier[linewidth=1.0pt](0.2,0)(0.2,0.5)(0.6,0.5)(0.6,1.0)
        \psbezier[linewidth=1.0pt](0.4,0)(0.4,0.5)(0.8,0.5)(0.8,1.0)
        \psbezier[linewidth=1.0pt](0.6,0)(0.6,0.2)(0.8,0.2)(0.8,0)
        \psbezier[linewidth=1.0pt](0.4,1.0)(0.4,0.8)(0.2,0.8)(0.2,1.0)
      }
      \endpspicture
      \:\right)   + \frac{d^2}{d^4-3d^2+2}
    \pspicture[0.4](0.9,0.9)
    \scalebox{0.9}{
      \psframe[linewidth=0.5pt](0,0)(1.0,1.0)
      \psbezier[linewidth=1.0pt](0.2,0)(0.2,0.2)(0.4,0.2)(0.4,0)
      \psbezier[linewidth=1.0pt](0.6,0)(0.6,0.2)(0.8,0.2)(0.8,0)
      \psbezier[linewidth=1.0pt](0.8,1.0)(0.8,0.8)(0.6,0.8)(0.6,1.0)
      \psbezier[linewidth=1.0pt](0.4,1.0)(0.4,0.8)(0.2,0.8)(0.2,1.0)
    }
    \endpspicture
    \\
    - \frac{d}{d^4-3d^2+2}
    \left(\: \pspicture[0.4](0.9,0.9)
      \scalebox{0.9}{
        \psframe[linewidth=0.5pt](0,0)(1.0,1.0)
        \psbezier[linewidth=1.0pt](0.2,0)(0.2,0.2)(0.4,0.2)(0.4,0)
        \psbezier[linewidth=1.0pt](0.6,0)(0.6,0.2)(0.8,0.2)(0.8,0)
        \psbezier[linewidth=1.0pt](0.8,1.0)(0.8,0.4)(0.2,0.4)(0.2,1.0)
        \psbezier[linewidth=1.0pt](0.6,1.0)(0.6,0.8)(0.4,0.8)(0.4,1.0)
      }
      \endpspicture
      + \pspicture[0.4](0.9,0.9)
      \scalebox{0.9}{
        \psframe[linewidth=0.5pt](0,0)(1.0,1.0)
        \psbezier[linewidth=1.0pt](0.2,0)(0.2,0.6)(0.8,0.6)(0.8,0)
        \psbezier[linewidth=1.0pt](0.4,0)(0.4,0.2)(0.6,0.2)(0.6,0)
        \psbezier[linewidth=1.0pt](0.8,1.0)(0.8,0.8)(0.6,0.8)(0.6,1.0)
        \psbezier[linewidth=1.0pt](0.4,1.0)(0.4,0.8)(0.2,0.8)(0.2,1.0)
}
\endpspicture
\:\right) + \frac{1}{d^4-3d^2+2}
\pspicture[0.4](0.9,0.9)
\scalebox{0.9}{
  \psframe[linewidth=0.5pt](0,0)(1.0,1.0)
  \psbezier[linewidth=1.0pt](0.2,0)(0.2,0.6)(0.8,0.6)(0.8,0)
  \psbezier[linewidth=1.0pt](0.4,0)(0.4,0.2)(0.6,0.2)(0.6,0)
  \psbezier[linewidth=1.0pt](0.8,1.0)(0.8,0.4)(0.2,0.4)(0.2,1.0)
  \psbezier[linewidth=1.0pt](0.6,1.0)(0.6,0.8)(0.4,0.8)(0.4,1.0)
}
\endpspicture
\\
=0,
\end{multline}
see Ref.~\onlinecite{Kauffman94} for a recursive formula. These
relations define a finite dimensional Hilbert space $V_d (Y) \subset
\overline{V}_d (Y) \subset \mathbb{C}^[S] \subset \mathbb{C}^S$.

In Refs.~\onlinecite{Freedman03,Good_guys} it is explained that these
$V_d(Y)$ are the Hilbert space for $\mathcal{D}K_k$ mentioned earlier,
$ d = 2\cos{\pi / (k+2)}$.  It is argued that a Hamiltonian with GSM
corresponding to $\overline{V}_d$ may be unstable and collapse under
perturbation (for $k=1$ or $2$ and under a larger deformation for
$k\geq 3$) to $V_d$.  Very briefly, TQFTs such as $\mathcal{D}K_k$,
can always be defined as the joint null space of commuting local
projectors\footnote{See Ref.~\onlinecite{Freedman01} where private
  communication with A.~Kitaev and G.~Kuperberg is referenced. The
  required family of commuting projectors is easily derived in the
  Turaev-Viro approach \cite{Turaev} by writing
  surface$\times$interval, $\Sigma \times I =$ handle-body union
  2-handles. The disjoint attaching curves of the 2-handles yield the
  commuting projectors.}, implying the existence of a local
Hamiltonian with a spectral gap in the thermodynamic limit.  Once a
Hamiltonian $H_d$ has imposed $d-$isotopy, i.e.  $GSM(H_d )=
\overline{V}_d$, an extensive degeneracy has been created; the only
local way of lifting this extensive degeneracy (to a finite
degeneracy) without creating frustration \footnote{In this paper we
  use the term ``frustration'' in reference to a Hamiltonian which can
  be written as a sum of projectors yet has no zero-modes. In this
  sense, new terms breaking the extensive degeneracy down to a
  crystalline structure will introduce frustration.} is to impose the
Jones-Wenzl projector as a constraint.  Although frustration may arise
in a Hamiltonian describing a topological phase, we know that these
phases can be produced by an unfrustrated Hamiltonian.  Thus it is an
attractive ansatz that near $H_d$ will be some $H_{d, \epsilon}$ with
$GSM(H_{\epsilon , d}) =V_d$. 

To this point our discussion has contemplated smooth multiloops $X$ on
a surface $Y$; now it is appropriate to move to a lattice setting. It
is an old idea (see e.g. Ref.~\onlinecite{Sutherland88}) to turn a
dimerization (perfect matching) $\mathcal{J}$ into a multiloop
$\mathcal{R} \cup \mathcal{J}$ by using a background dimerization
$\mathcal{R}$ to form a `transition graph'.  If the lattice
$\mathcal{L}$ is bipartite then $\mathcal{R} \cup \mathcal{J}$ can be
oriented in a natural way, leading to conserved quantities
\footnote{If a lattice is bipartite, i.e. $\mathcal{L} = \mathcal{L}_A
  \cup \mathcal{L}_B$ so that for any site $a \in \mathcal{L}_A$ all
  of its neighbouring sites $b \in \mathcal{L}_B$ and vice versa, then
  any dimer $r \in \mathcal{R}$ can be oriented: $r\equiv (a,b) =
  (a\leftarrow b)$ with the opposite rule for dimers $j \in
  \mathcal{J}$: $j\equiv (a,b) = (a\to b)$.  Then the transition graph
  $\mathcal{R} \cup \mathcal{J}$ consists of oriented loops
  (orientation of loops of length 2 is ambiguous but can be ignored --
  it is the essential loops that lead to conserved quantities). From
  the physical point of view, these oriented loops can be treated as
  ``level sets'' or ``equipotentials'' leading to the height
  representation with the conserved quantity being the overall tilt.
  Such conserved quantities severely restrict quantum dynamics and
  typically lead to either crystalline phases or gapless quantum
  critical points \cite{Moessner03:SMA}.} -- integral winding numbers
-- which are disrespected by the Jones-Wenzl (JW) relations.  For this
reason we pass over the square and hexagonal lattices as poor
candidates for the imposition of the JW relations crucial to the
passage from $\overline{V}_d$ to $V_d$.  Thus we consider perfect
matchings on the triangular lattice $\mathcal{T}$.  By fixing
$\mathcal{R}$ as in Fig.~\ref{fig:Triang}, without small rhombi with
two opposite sides red, as the preferred background dimerization we
obtain the fewest equations in defining $\overline{V}_d$ and also
achieve ergodicity \footnote{D.~Jetchev, private communication. For a
  rigorous proof of ergodicity in a similar setting see
  Ref.~\onlinecite{Kenyon96}.} under a small set of moves.  Unlike in
the usual case, the the background dimerization, $\mathcal{R}$, is not
merely a guide for the eyes, but will be \emph{physically}
distinguished: the chemical potentials and tunnelling amplitudes will
be different for bonds of different color.

\begin{figure}[htb]
\includegraphics[width=2.75in]{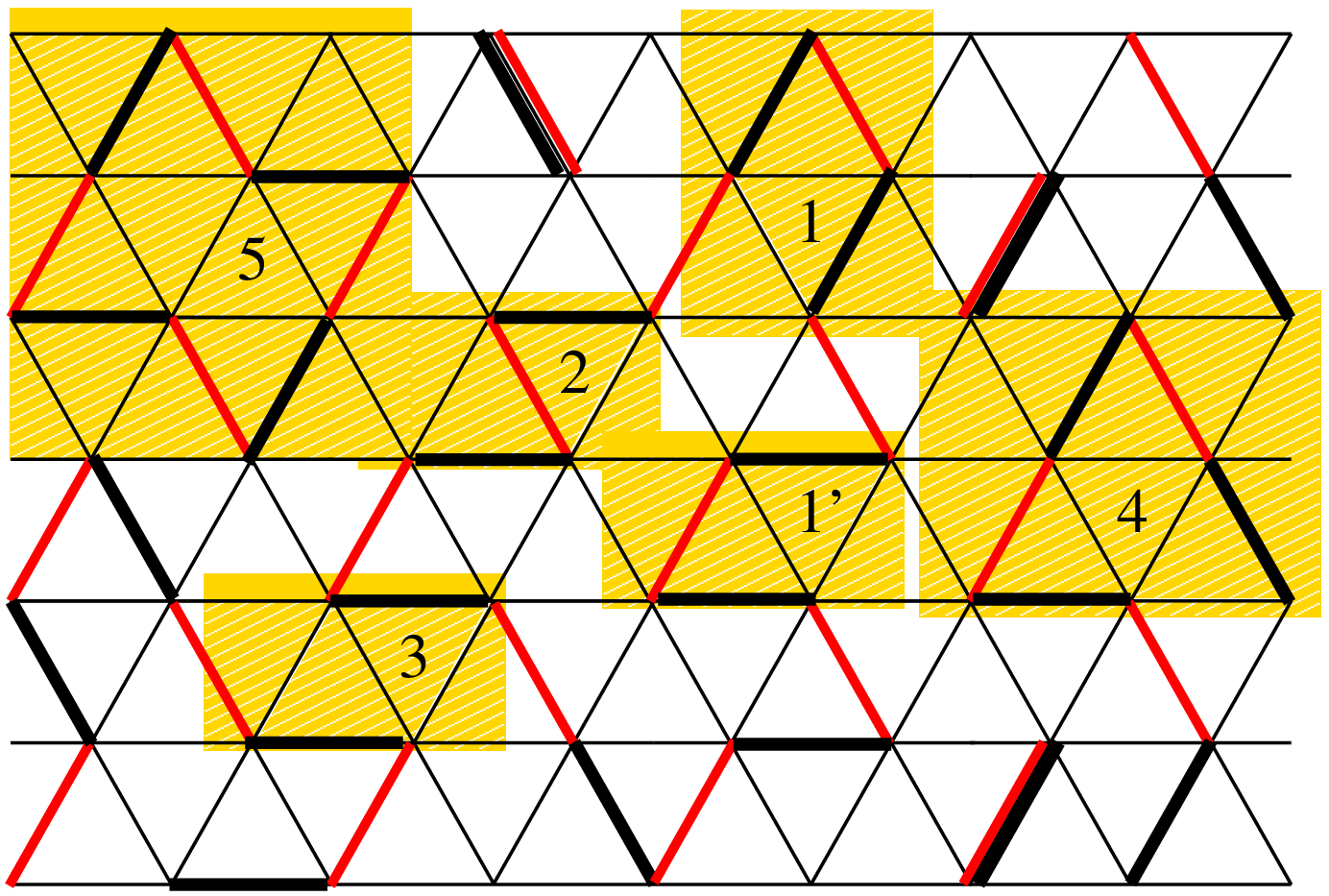}
\caption{
  Overlap of a dimer covering of $\mathcal{T}$ (shown in thick black)
  with the red covering corresponding to the special sublattice
  $\mathcal{R}$.  Shaded plaquettes correspond to various dimer moves
  described in the text. Green sublattice is not shown.}
\label{fig:Overlap}
\end{figure}

Let us list here the elementary dimer moves that preserve the proper
dimer covering condition:
\begin{itemize}
\item{Plaquette (rhombus) flip -- this is a two-dimer move around a
    rhombus made of two lattice triangles. Depending on whether a
    ``red'' bond forms a side of such a rhombus, its diagonal, or is
    not found there at all, the plaquettes are referred to,
    respectively, as type 1 (or 1'), 2, or 3 (see
    Fig.~\ref{fig:Overlap}). The distinction between plaquettes of
    type 1 and 1' is purely directional: diagonal bonds in plaquettes
    of type 1 are horizontal, for type 1' they are not. This
    distinction is necessary since our Hamiltonian breaks the
    rotational symmetry of a triangular (or Kagom\'{e}) lattice.}
\item{Triangle move -- this is a three-dimer move around a triangle
    made of four elementary triangles. One such ``flippable'' triangle
    is labelled 4 in Fig.~\ref{fig:Overlap}.}
\item{Bow tie move -- this is a four-dimer move around a ``bow tie''
    made of six elementary triangles. One such ``flippable'' bow tie
    is labelled 5 in Fig.~\ref{fig:Overlap}.}
\end{itemize}
To make each of the above moves possible, the actual dimers and
unoccupied bonds should alternate around a corresponding shape.
Notice that for both triangle and bow tie moves we chose to depict the
cases when the maximal possible number of ``red'' bonds participate in
their making (2 and 4 respectively).

Note that there are no alternating red/black rings of fewer than 8 lattice
bonds (occupied by at most 4 non-colliding dimers). Ring moves only occur
when red and black dimers alternate; the triangle labelled 4 in 
Fig.~\ref{fig:Overlap} does not have a Ring term associated with it,
but the bow tie labelled 5 does.

Here is the correspondence between the previous smooth discussion and
rhombus flips relating dimerizations $\mathcal{J}$ of $\mathcal{T}$.
The surface $Y$ is now a planar domain or possibly a torus (periodic
boundary conditions). A multiloop $X$ in $Y$ becomes $X=\mathcal{R}
\cup \mathcal{J}$ (with the convention that the dimers of $\mathcal{R}
\cap \mathcal{J}$ be consider as length 2 loops or bigons). What about
isotopy?  Move (2) certainly is an isotopy from $\mathcal{R} \cup
\mathcal{J}$ to $\mathcal{R} \cup \mathcal{J}'$ but move (2), by
itself, does almost nothing.  It is not possible to build up long
motions from (2) alone.  So it is a peculiarity of the rhombus flips
that we have no good analog of isotopy alone but instead go directly
to $d-$isotopy. Move (1) must be considered in two different forms (1)
and (1') as a result of the color differences (see
Fig.~\ref{fig:Overlap}).  The reader may expect us to impose the
relations associated with type (5) and (1), (1') moves:
\begin{subequations}
  \label{dimer_rel_1}
  \begin{eqnarray}
    & & d^3\; \Psi \Bigg(
    \pspicture[-0.05](.8,0.69282)
    \psset{linewidth=2pt}
    \psline[linecolor=black](0,0.34641)(0.4,0.34641)
    \psline[linecolor=red](0.4,0.34641)(0.8,0.34641)
    \psline[linecolor=black](0.8,0.34641)(0.6,0)
    \psline[linecolor=red](0.6,0)(0.8,-0.34641)
    \psline[linecolor=black](0.8,-0.34641)(0.4,-0.34641)
    \psline[linecolor=red](0.4,-0.34641)(0,-0.34641)
    \psline[linecolor=black](0,-0.34641)(0.2,0)
    \psline[linecolor=red](0.2,0)(0,0.34641)
    \psline[linecolor=black,linestyle=dotted,linewidth=1pt](0.2,0)(0.6,0)
    \pspolygon[linecolor=black,linestyle=dotted,linewidth=1pt](0.2,0)
    (0.4,0.34641)(0.6,0)(0.4,-0.34641)
    \endpspicture
    \Bigg)
    - \Psi \Bigg(
    \pspicture[-0.05](.8,0.69282)
    \psset{linewidth=1.5pt}
    \psline[linecolor=black,linestyle=dotted,linewidth=1pt](0,0.34641)
    (0.4,0.34641)
    \psline[linecolor=red](0.4,0.34641)(0.8,0.34641)
    \psline[linecolor=black](0.4,0.38641)(0.8,0.38641)
    \psline[linecolor=black,linestyle=dotted,linewidth=1pt](0.8,0.34641)
    (0.6,0)
    \psline[linecolor=red](0.6,0)(0.8,-0.34641)
    \psline[linecolor=black](0.63,0.03)(0.83,-0.31641)
    \psline[linecolor=black,linestyle=dotted,linewidth=1pt](0.8,-0.34641)
    (0.4,-0.34641)
    \psline[linecolor=red](0.4,-0.34641)(0,-0.34641)
    \psline[linecolor=black](0.4,-0.30641)(0,-0.30641)
    \psline[linecolor=black,linestyle=dotted,linewidth=1pt](0,-0.34641)
    (0.2,0)
    \psline[linecolor=red](0.2,0)(0,0.34641)
    \psline[linecolor=black](0.23,0.03)(0.03,0.37641)
    \psline[linecolor=black,linestyle=dotted,linewidth=1pt](0.2,0)(0.6,0)
    \pspolygon[linecolor=black,linestyle=dotted,linewidth=1pt](0.2,0)
    (0.4,0.34641)(0.6,0)(0.4,-0.34641)
    \endpspicture
    \Bigg)
    =0,
    \label{dimer_rel_1a}
    \\
    & & d\; \Psi \left(
    \pspicture[0.0](.6,0.34641)
      \psset{linewidth=2pt}
      \psline[linecolor=black](0,-0.1)(0.2,0.24641)
      \psline[linecolor=red](0.2,0.24641)(0.6,0.24641)
      \psline[linecolor=black](0.6,0.24641)(0.4,-0.1)
      \psline[linecolor=black,linestyle=dotted,linewidth=1pt](0,-0.1)(0.4,-0.1)
      \psline[linecolor=black,linestyle=dotted,linewidth=1pt](0.2,0.24641)
      (0.4,-0.1)
      \endpspicture
    \right)
    - \Psi \left(
      \pspicture[0.0](.6,0.34641)
      \psset{linewidth=2pt}
      \psline[linecolor=black,linestyle=dotted,linewidth=1pt](0,-0.1)
      (0.2,0.24641)
      \psline[linecolor=red,linewidth=1.5pt](0.2,0.24641)(0.6,0.24641)
      \psline[linecolor=black,linewidth=1.5pt](0.2,0.28641)(0.6,0.28641)
      \psline[linecolor=black,linestyle=dotted,linewidth=1pt](0.6,0.24641)
      (0.4,-0.1)
      \psline[linecolor=black](0,-0.1)(0.4,-0.1)
      \psline[linecolor=black,linestyle=dotted,linewidth=1pt](0.2,0.24641)
      (0.4,-0.1)
      \endpspicture
    \right)
    =0
    \label{dimer_rel_1b}
  \end{eqnarray}
\end{subequations}
since we pass from one to four loops in (\ref{dimer_rel_1a}) and zero
to one loop in (\ref{dimer_rel_1b}).

However, by choosing, instead a less obvious mapping of
$\overline{V}_d$ to functions on $\{\mathcal{J}\}$ we end up imposing
one fewer equation on the Hubbard parameters. So instead we impose:
\begin{subequations}
  \label{dimer_rel_2}
  \begin{eqnarray}
    & & b\; \Psi \Bigg(
    \pspicture[-0.05](.8,0.69282)
    \psset{linewidth=2pt}
    \psline[linecolor=black](0,0.34641)(0.4,0.34641)
    \psline[linecolor=red](0.4,0.34641)(0.8,0.34641)
    \psline[linecolor=black](0.8,0.34641)(0.6,0)
    \psline[linecolor=red](0.6,0)(0.8,-0.34641)
    \psline[linecolor=black](0.8,-0.34641)(0.4,-0.34641)
    \psline[linecolor=red](0.4,-0.34641)(0,-0.34641)
    \psline[linecolor=black](0,-0.34641)(0.2,0)
    \psline[linecolor=red](0.2,0)(0,0.34641)
    \psline[linecolor=black,linestyle=dotted,linewidth=1pt](0.2,0)(0.6,0)
    \pspolygon[linecolor=black,linestyle=dotted,linewidth=1pt](0.2,0)
    (0.4,0.34641)(0.6,0)(0.4,-0.34641)
    \endpspicture
    \Bigg)
    - \Psi \Bigg(
    \pspicture[-0.05](.8,0.69282)
    \psset{linewidth=1.5pt}
    \psline[linecolor=black,linestyle=dotted,linewidth=1pt](0,0.34641)
    (0.4,0.34641)
    \psline[linecolor=red](0.4,0.34641)(0.8,0.34641)
    \psline[linecolor=black](0.4,0.38641)(0.8,0.38641)
    \psline[linecolor=black,linestyle=dotted,linewidth=1pt](0.8,0.34641)
    (0.6,0)
    \psline[linecolor=red](0.6,0)(0.8,-0.34641)
    \psline[linecolor=black](0.63,0.03)(0.83,-0.31641)
    \psline[linecolor=black,linestyle=dotted,linewidth=1pt](0.8,-0.34641)
    (0.4,-0.34641)
    \psline[linecolor=red](0.4,-0.34641)(0,-0.34641)
    \psline[linecolor=black](0.4,-0.30641)(0,-0.30641)
    \psline[linecolor=black,linestyle=dotted,linewidth=1pt](0,-0.34641)
    (0.2,0)
    \psline[linecolor=red](0.2,0)(0,0.34641)
    \psline[linecolor=black](0.23,0.03)(0.03,0.37641)
    \psline[linecolor=black,linestyle=dotted,linewidth=1pt](0.2,0)(0.6,0)
    \pspolygon[linecolor=black,linestyle=dotted,linewidth=1pt](0.2,0)
    (0.4,0.34641)(0.6,0)(0.4,-0.34641)
    \endpspicture
    \Bigg)
    =0,
    \label{dimer_rel_2a}
    \\
    & & a\; \Psi \left(
    \pspicture[0.0](.6,0.34641)
      \psset{linewidth=2pt}
      \psline[linecolor=black](0,-0.1)(0.2,0.24641)
      \psline[linecolor=red](0.2,0.24641)(0.6,0.24641)
      \psline[linecolor=black](0.6,0.24641)(0.4,-0.1)
      \psline[linecolor=black,linestyle=dotted,linewidth=1pt](0,-0.1)(0.4,-0.1)
      \psline[linecolor=black,linestyle=dotted,linewidth=1pt](0.2,0.24641)
      (0.4,-0.1)
      \endpspicture
    \right)
    - \Psi \left(
      \pspicture[0.0](.6,0.34641)
      \psset{linewidth=2pt}
      \psline[linecolor=black,linestyle=dotted,linewidth=1pt](0,-0.1)
      (0.2,0.24641)
      \psline[linecolor=red,linewidth=1.5pt](0.2,0.24641)(0.6,0.24641)
      \psline[linecolor=black,linewidth=1.5pt](0.2,0.28641)(0.6,0.28641)
      \psline[linecolor=black,linestyle=dotted,linewidth=1pt](0.6,0.24641)
      (0.4,-0.1)
      \psline[linecolor=black](0,-0.1)(0.4,-0.1)
      \psline[linecolor=black,linestyle=dotted,linewidth=1pt](0.2,0.24641)
      (0.4,-0.1)
      \endpspicture
    \right)
    =0
    \label{dimer_rel_2b}
  \end{eqnarray}
\end{subequations}
and require that ${a^4}/{b}=d$.

In other words, the two processes, (1) and (1') for making length 2
loops and the one process (5) for fusing four length 2 loops in one loop
of length 8 (see Fig.~\ref{fig:Overlap}) will have to be tuned to
satisfy Eqs.~(\ref{dimer_rel_2}) and ${a^4}/{b}=d$.

Given $\Psi \in \overline{V}_d$, let $\Psi_{\rho}(\mathcal{J}) =
a^{\#} \Psi(( \mathcal{R} \cup \mathcal{J} )^-)$ where $\#$ is the
number of length 2 loops of $\mathcal{R} \cup \mathcal{J}$ (i.e.  $\#$
of dimers common to $\mathcal{R}$ and $\mathcal{J}$) and $(\mathcal{R}
\cup \mathcal{J})^-$ is the multiloop $\mathcal{R} \cup \mathcal{J}
\setminus \mathcal{R} \cap \mathcal{J}$, i.e. all of the loops except
the length 2 loops. This pulls $\Psi$, a function on multiloops back
to a function $\Psi_{\rho}(\mathcal{J})$ on dimerizations.  If
$\mathcal{J}$ and $\mathcal{J}'$ are linked by finitely many
applications of Eqs.~(\ref{dimer_rel_2}) and the isotopy move, it is
now easy to check that $\Psi_{\rho}(\mathcal{J}) = a^{(\#' -\#)}
d^{(c' -c)} \Psi_{\rho}(\mathcal{J}')$ where $c(\text{resp.  }c')$ is
the number of non-essential loops of length exceeding two in
$\mathcal{R} \cup \mathcal{J} ( \text{ resp. } \mathcal{R} \cup
\mathcal{J}' )$.

This separate accounting for length 2 loops and the longer circles may
appear to be a slight of hand.  Is there a price to
pay? In a sense, yes: It is now crucial that a combinatorial relation
which mimics the smooth JW relation by relating $(\mathcal{J}_1 ,
\mathcal{J}_2 )$ for $k=1$, $(\mathcal{J}_1, \ldots, \mathcal{J}_5 )$
for level $2$, $(\mathcal{J}_1 , \ldots \mathcal{J}_{14} )$ for level
3, etc$\ldots$ must not in any of its terms change the number of
length 2 loops.  But this is just an additional (and readily achieved)
condition on the combinatorial form of the JW projector and does not
influence the algebraic conclusion: that (only) for special $d$ is
there a unfrustrated local reduction $V_d \subset \overline{V}_d$
which could be a stable phase and when $d$ is special $V_d$ is unique.
Thus the combinatorial analog of $\overline{V}_d$ is functions on
$\{\mathcal{J}\}$ obeying Eqs.~(\ref{dimer_rel_2}) with ${a^4}/{b}=d$.

\section{The analysis of $A$ and $A^-$}
\label{sec:model_A}

In this section we derive the effective Hamiltonian $\tilde{H}:
\mathcal{D}\to \mathcal{D}$ on the span of dimerizations. The
derivation is perturbative to the second order in $\epsilon$ where
$\epsilon=t^{r}_{bb}/U=t^{b}_{gb}/U$. Additionally, $t^{b}_{rb}/U =
c_0 \epsilon$ where $c_0$ is a positive constant, while
$t^{g}_{bb}={o}(\epsilon)$ and can be neglected in the second-order
calculations. (In the absence of a magnetic field all $t$'s can be
made real and hence symmetric with respect to their lower indices.
Also, from now on we set $U =1$ for notational convenience.)

As explained in Section~\ref{sec:d-isotopy}, we then shall tune
$\tilde{H}$ to the ``small loop'' value $a$ and the ``bow tie'' value
$b$ with $a^4 /b = d$.

We account for all second-order processes, i.e. those processes that
take us out of $\mathcal{D}$ and then back to $\mathcal{D}$ (see
Appendix~\ref{appendix2} for technical details). As mentioned earlier,
these amount to off-diagonal (hopping) processes -- ``plaquette''
flips'' or ``rhombus moves'' -- as well as diagonal ones (potential
energy) in which a dimer pivots out and then back into its original
position.  The latter processes lead to renormalisation of the bare
onsite potentials $\mu_i$, which we have adjusted so that all
renormalised potentials $\tilde\mu_i$ are equal up to corrections
$\mathcal{O}(\epsilon^3)$. The non-constant part of the effective
Hamiltonian comes from the former processes and can be written in the
form:
\begin{equation}
  \label{eq:eff_form}
  \tilde{H}= \sum_{\mathcal{I},\mathcal{J}}
  \left( \tilde{H}_{\mathcal{I} \mathcal{J}} \otimes \mathbb{I}\right) 
  \tilde \delta_{\mathcal{I} \mathcal{J}}
\end{equation}
where $\tilde{H}_{\mathcal{I}\mathcal{J}}$ is a $2\times 2$ matrix
corresponding to a dimer move in the two-dimensional basis of dimer
configurations connected by this move. $\tilde \delta_{\mathcal{I}
  \mathcal{J}}=1$ if the dimerizations ${\mathcal{I},\mathcal{J}}\in
\mathcal{D}$ are connected by an allowed move, $\tilde
\delta_{\mathcal{I} \mathcal{J}}=0$ otherwise.

Therefore it suffices to specify these $2\times 2$ matrices
$\tilde{H}_{\mathcal{I} \mathcal{J}}$ for the off-diagonal processes.
For moves of types (1)-(3), these matrices are given below.
Type (1), e.g. rhombus $(VY)$ in Fig.~\ref{fig:model_B}:
\begin{multline}
  \label{eq:move_1}
  \begin{pmatrix}
    v^{b}_{gb}&  -2 t^{b}_{rb} t^{b}_{gb}    \\
    -2 t^{b}_{rb} t^{b}_{gb}&  v^{b}_{rb}
  \end{pmatrix}
  =
  \begin{pmatrix}
    v^{b}_{gb}&  -2 c_0 \epsilon^{2}    \\
    -2 c_0 \epsilon^{2}&  v^{b}_{rb}
  \end{pmatrix}
  \\
  \sim
  \begin{pmatrix}
    a &  -1  \\
    -1 &  a^{-1}
  \end{pmatrix}
\end{multline}
Type (1'), e.g. rhombus $(VW)$ in Fig.~\ref{fig:model_B}:
\begin{multline}
\label{eq:move_1_prime}
  \begin{pmatrix}
    v^{b}_{bb}&  -2 t^{b}_{rb} t^{b}_{gb}    \\
    -2 t^{b}_{rb} t^{b}_{gb}&  v^{b}_{rg}
  \end{pmatrix}
  =
  \begin{pmatrix}
    v^{b}_{bb}&  -2 c_0 \epsilon^2    \\
    -2 c_0 \epsilon^2 &  v^{b}_{rb}
  \end{pmatrix}
  \\
  \sim
  \begin{pmatrix}
    a &  -1  \\
    -1 &  a^{-1}
  \end{pmatrix}
\end{multline}
Type (2), e.g. rhombus $(UV)$ in Fig.~\ref{fig:model_B}:
\begin{multline}
\label{eq:move_2}
  \begin{pmatrix}
   v^{r}_{bb}&  -2 (t^{r}_{bb})^{2}    \\
   -2 (t^{r}_{bb})^{2} &  v^{r}_{bb}
  \end{pmatrix}
  =
 \begin{pmatrix}
   v^{r}_{bb}&  -2 \epsilon^{2}   \\
    -2 \epsilon^{2}&  v^{r}_{bb}
  \end{pmatrix}
  \\
  \sim
  \begin{pmatrix}
    1 &  -1  \\
    -1 &  1
  \end{pmatrix}
\end{multline}
since isotopic multiloops should have equal weights.

Finally, type (3), e.g. rhombus $(WX)$ in Fig.~\ref{fig:model_B}:
\begin{multline}
\label{eq:move_3}
  \begin{pmatrix}
    v^{g}_{bb}&  -2 (t^{g}_{bb})^{2}    \\
    -2 (t^{g}_{bb})^{2} &  v^{g}_{bb}
  \end{pmatrix} 
  =
  \begin{pmatrix}
    v^{g}_{bb} &  0   \\
    0 & v^{g}_{bb}
  \end{pmatrix}
  =
  \begin{pmatrix}
    0&  0   \\
    0& 0
  \end{pmatrix},
\end{multline}
provided $k>1$,
since it represents a ``surgery'' similar to Eq.~(\ref{eq:JW1}).
(For $k~=~1$, on the other hand, this matrix must be proportional
to $\left( \begin{smallmatrix}
    1&  -1   \\
    -1& 1
  \end{smallmatrix}\right)$.)
As follows from Eq.~(\ref{eq:JW1}), at level $k=1$ multiloops which
differ by a surgery should have equal coefficients in any ground state
vector $\Psi$ while at levels $k>1$ no such relation should be
imposed. We use the symbol ``$\sim$'' to denote proportional via a
positive factor.

The matrix relations (\ref{eq:move_1}-\ref{eq:move_3}) yield equations
in the model parameters (we consider now only the case $k>1$):
\begin{subequations}
\label{eq:param_cond_A}
\begin{eqnarray}
  \text{Types } (1) \& (1'):  &  v^b_{gb} = v^b_{bb} = 2ac_0 \epsilon^2 
  \\
  { } & \text{and}\quad v^b_{rb} = v^b_{rg}= 2a^{-1} c_0 \epsilon^2
  \\
  \text{Type } (2):  & v^r_{bb} = 2\epsilon^2 \\
  \text{Type } (3): & v^g_{bb} = 0
\end{eqnarray}
\end{subequations}

We suppose that the Hamiltonian has a bare ring exchange term,
Ring in Eq.~(\ref{eq:Hubbard}): 
$\text{Ring}= \left(
  \begin{smallmatrix}
    x  &  -c_3 \epsilon^2 \\
    -c_3 \epsilon^2  &  y
  \end{smallmatrix}
\right) $ for some constants $c_3,\; x,\; y>0$, and consider the
additional equations which come from considering Ring as a fluctuation
between one loop of length 8 (labelled (5) in Fig.~\ref{fig:Overlap})
and four loops of length 2 .  It follows from Eq.~(\ref{dimer_rel_2a})
that $\text{Ring} \sim \left(
  \begin{smallmatrix}
    b  &  -1 \\
    -1  &  1/b
  \end{smallmatrix}
\right)
$, 
$b>0$ so:
\begin{multline}
  \label{eq:3.1}
  {a^4}/{b} = d,
  \quad
  {x}/{\epsilon^2}  = b c_3,
  \quad
  {y}/{\epsilon^2} = b^{-1} c_3 .
\end{multline}
are the additional equations (beyond Eqs.~(\ref{eq:param_cond_A})) to
place model $A$ at the soluble point $\overline{V}_d $.  The
justification of the diagonal entries in the ring term is that the 4
particles in $\Psi_0$ and $\Psi_1$ form a square which has some cost
over an ideal Winger crystal of particles.  This cost can be
influenced by the local chemical environment so the entries do not
need to be equal, though $x=y$ is the most natural case.

Computer studies \cite{Jetchev:priv} show that with the Ring term
present all $\mathcal{J}$ are connected by repeated application of the
Hamiltonian (and respect the $d-$isotopy relation).  When the Ring
term is removed, it still appears that any two dimerizations
$\mathcal{J}$ and $\mathcal{J}'$ which determine isotopic loops
nesting patterns $(\mathcal{R} \cup \mathcal{J})$ and
$(\mathcal{R}\cup \mathcal{J}')$ (use our conversion to consider
bigons as loops) communicate. So, e.g., in a disk if $\mathcal{R}\cup
\mathcal{J}$ and $\mathcal{R}\cup \mathcal{J}'$ have the same number
of loops and the same combinatorial nesting relations, and if
$\mathcal{R}$, $\mathcal{J}$, and $\mathcal{J}'$ all agree near the
disk boundary, then moves of type (1), (1'), (2), and their reversals
will connect $\mathcal{J}$ and $\mathcal{J}'$. Since the Jones-Wenzl
relations do violence to the nesting patterns it is hard to imagine
any \emph{additional} conserved quantities which could allow the
ground state \emph{without} the Ring term to collapse to a richer
(move degenerate) topological phase than $\mathcal{D}K_k$ itself (for
some $k$).  So we suspect that the Ring term is redundant if the goal
is to arrive in a topological phase.

However, there is a caution which should be issued if the ring term is
omitted.  Because of the possibility of defining a separate circle
value $a$ for bigons and a value $b$ for breaking 8-gons, with only
the relation $\frac{a^4}{b} = d$ tying the model to a fixed
level{\footnote{Recall that replacing the constant $d$ with
    variables $a$ and $b$ adds an important degree of freedom to the
    model.} the $A^-$ model (Eq.~(\ref{eq:Hubbard}) with Ring omitted)
  could still be tuned to any level $k$ because only $a$ and not $b$
  has been given.  Thus, for example $\mathcal{D}K_1$ can still
  arise if an effective Ring term with $b=a^4$ emerges.
  
  But, as we have remarked, physically it is most natural to assume
  $x=y$ in the Ring term. This forces $b=1$ and $a=d^{{1}/{4}}$.  Note
  that we do \emph{not} want fluctuations on the alternating
  green-black $8-$bond rings as this would mix distinct topological
  sectors and no such terms are in the Hamiltonian (\ref{eq:Hubbard}).
  
  Our simplest bosonic candidate for a ``universal quantum computer''
  would be model $A$ tuned to
  $a=\left(\frac{1+\sqrt{5}}{2}\right)^{1/4}$; $x=y = c_3 \epsilon^2$
  in Ring (and no green-black Ring term).

\section{ Model $B$ (spinless boson/no Ring term)}

For this model we work entirely within the extended Hubbard model,
$H$, as in Eq.~(\ref{eq:Hubbard}) with no Ring term.  What replaces
the 4-particle ring exchange term is a flip of a new rhombus of type
$(r,r)$: $\pspicture[0.0](.6,0.34641) \psset{linewidth=2pt}
\psline[linecolor=black](0,-0.1)(0.2,0.24641)
\psline[linecolor=red](0.2,0.24641)(0.6,0.24641)
\psline[linecolor=black](0.6,0.24641)(0.4,-0.1)
\psline[linecolor=red](0,-0.1)(0.4,-0.1)
\psline[linecolor=black,linestyle=dotted,linewidth=1pt](0.2,0.24641)
(0.4,-0.1) \endpspicture $
(e.g. rhombus $QP$ in Fig.~\ref{fig:model_B}) which can be interpreted
as an alternating ring of length 4.  We create an extensive system of
(for example, bi-periodic) ``defects'' in the red sublattice
$\mathcal{R}$.  A defect is made by rotating four red edges by a
${1}/{8}$-turn around a bow tie.  Inspecting this defect reveals two
$(r, r)$-rhombi adjacent to the defect.  The required $(r,r)$-rhombus
flip can easily be coaxed out of the Hubbard model at $\mathcal{O}
(\epsilon^2)$ so there is no need to include an ad hoc Ring term.
(Without a defect a Ring term does not arise until order $\mathcal{O}
(\epsilon^4)$ -- see Appendix~\ref{appendix2}.)

We can construct a $d-$isotopy space $\overline{V}_d$ as the GSM of
model $B$ in a similar fashion to model $A^-$, but the price of having
introduced the defect is that there are now many more exceptional
cases for type (1), (2), and (3) moves.  A proliferation of colors
must be defined so that raw potentials $\mu_i$ when renormalized by
hopping in all the different local environments come out equal up to
$\mathcal{O} (\epsilon^2)$.  We introduce the colors (see
Fig.~\ref{fig:model_B}): $g$(green), $b$(black), $i$(indigo),
$r$(red), $l$(lavender), and $G$(thick green), $B$(thick black) which
are identical in tunnelling properties to $g$ and $b$ (resp.) but
require different on-site potentials $\mu_{G}\neq \mu_{g}$ and $
\mu_{B}\neq \mu_{b}$.

\begin{figure}[hbt]
  \includegraphics[width=2.75in]{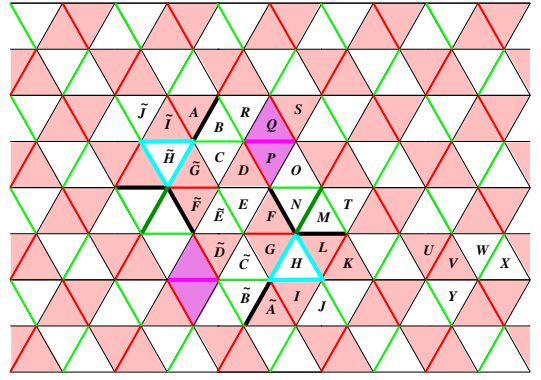}
  \caption{Sublattice defect, a single bow tie rotation of 
    $\mathcal{R}$, in bosonic model $B$. Two rhombi 
    with alternating red and black edges (type $(r,r)$) are shaded
    lavender. Triangles forming plaquettes whose color environment 
    is affected by the defect are labelled $A-T$ and
    $\tilde{A}-\tilde{T}$ (with notations reflecting the obvious
    inversion symmetry). Triangles labelled $U-Y$ form regular
    plaquettes.}
  \label{fig:model_B}
\end{figure}

We take the following relations (which are not the most general
possible):
\begin{subequations}
  \label{eq:rules_B_t}
  \begin{eqnarray}
    &  &  0 = t^{g}_{bb} = t^{g}_{gb}=t^{G}_{gb}=t^{g}_{Gb} \\
    &  &  \epsilon = t^{r}_{bb} = t^{b}_{gb}=t^{r}_{bl}=t^{i}_{ii}
    = t^{r}_{bi} = t^{b}_{gg}=t^{b}_{gG}\\
    &  &  c_0 \epsilon = t^{b}_{lr} = t^{b}_{ir}= t^{x}_{br}
    =t^{l}_{bb},\quad \text{for } x \neq l \\
    &  &   c_1 \epsilon = t^{l}_{br}
  \end{eqnarray}
\end{subequations}
(Note: With regard to tunnelling amplitudes, $i$ behaves like $b$
except $t^{i}_{ii} = \epsilon$ which has no $b$--analog.)

Near the defect in $\mathcal{R}$ many special rhombi occur and must be
considered. All labelled rhombi (those bearing two letters in
Fig.~\ref{fig:model_B}) can be classified by the topological effect of
a two-dimer move: types (1) and (1') create or absorb a loop of length
2, type (2) is an isotopy move, type (3) is a surgery and an
additional type $(r,r)$ (e.g. rhombus $(QP)$) breaks a 4-gon into two
bigons (or the reverse). Regard the interaction potentials $v$ as
variables and for each labelled rhombus write the matrix relations,
extending Eqs.~(\ref{eq:move_1})--(\ref{eq:move_3}). we may now solve
for the $v$'s as functions of the $t$'s as shown below
(\ref{eq:rules_B_v}-\ref{eq:rules_B_v_add}). To shorten the list of
$v$'s we assume that capitalisation of indices for $v$ has no effect.
Similarly, we assume that $i$ and $b$ are interchangeable.
\begin{subequations}
  \label{eq:rules_B_v}
  \begin{eqnarray}
    \text{Type } (3): & 0 = v^{g}_{bb} = v^{g}_{bg} \\
    \text{Type } (2):  & 2\epsilon^2 = v^{r}_{bb} = v^{r}_{bl} \\
    \text{Types } (1) \& (1'):  
    & 2a c_0 \epsilon^2 = v^{b}_{gb} = v^{b}_{bb}
    = v^{b}_{bi}=v^{b}_{lb} \\
    & \text{and}\quad 2a^{-1}c_0 \epsilon^2 = v^{b}_{rb} = v^{b}_{rg}
  \end{eqnarray}
\end{subequations}

The only relation not previously appearing in model $A^-$ comes from
from type $(r,r)$ rhombi (e.g. $(QP)$): 
\begin{multline}
  \label{eq:rules_B_v_add}
  \left(\begin{matrix}
      v^{l}_{bb}& -2(t^{l}_{rb})^2   \\
      -2(t^{l}_{rb})^2 & v^{l}_{rr}
    \end{matrix}\right)
  \sim 
  \left(\begin{matrix}
      b & -1  \\
      -1 & b^{-1}
    \end{matrix}\right)
  \\
  \Rightarrow 
  v^{l}_{bb}  =  2b c^2_1 \epsilon^2 ,\,\,
  v^{l}_{rr}  =  2 b^{-1} c^2_1 \epsilon^2,
\end{multline}
with the additional constraint $ {a^2}/{b} =d$.

In principle, all labelled rhombi give relations on $v$'s as functions
of the $t$'s, but we have set things up so that these further
relations are redundant with the ones coming from $VY$, $VW$, $UV$,
$WX$ and $QP$ which we have already used in deriving
(\ref{eq:rules_B_v}-\ref{eq:rules_B_v_add}). In the following table we
begin a verification which the reader may complete.

Type (1'), rhombus $(\tilde{J}\tilde{I})$:
  \begin{equation}
    \begin{pmatrix}
      v^{b}_{ib} & -2 t^{b}_{gb}t^{b}_{ri}\\
      -2t^b_{gb} t^b_{ri} & v^b_{rg}
    \end{pmatrix}
    \;\;
    \Rightarrow 
    \;\;
    t^{b}_{ri}= c_0 \epsilon ,\:\: v^i_{ib} = v^b_{bb}
  \end{equation}
  Type (2), rhombus $(\tilde{I}A)$:
  \begin{multline}  
    \begin{pmatrix}
      v^{r}_{ib} & -2(t^{r}_{bb})^2\\
      -2(t^{r}_{bb})^2& v^{r}_{Bb}
    \end{pmatrix}
    \\
    \Rightarrow v^r_{ib} =v^r_{bb}, \,\, v^r_{Bb} = v^h_{bb},\,\,
    t^r_{bb} = \epsilon
      \end{multline}
      Type (1'), rhombus $(AB)$:
  \begin{multline}  
    \begin{pmatrix}
      v^{B}_{bg}& -2t^{B}_{gg} t^B_{rb}  \\
      -2t^{B}_{gg} t^B_{rb}& v^{B}_{rg}
    \end{pmatrix}
    \\
    \Rightarrow t^B_{rb} = c_0 \epsilon ,\,\, t^B_{gg} =\epsilon ,\,\,
    v^B_{bg}=v^b_{bg}, \,\, v^B_{gr}=v^b_{gr}
  \end{multline}
\section{Fermionic Models $C$ and $D$}
For $C$ the Hamiltonian is again (\ref{eq:Hubbard}) but with $c_i$ and
$c^\dagger_1$ spinless fermionic annihilation and creation and
operators.  The lattice will be triangular $\mathcal{T}$.  The special
background dimerization will be the red sublattice $\mathcal{R}$.  The
green sublattice $\mathcal{G}$ is also marked and edges of
$\mathcal{B}:= \mathcal{T} \setminus \left(\mathcal{R}\cup \mathcal{G}
\right)$ will be called black, see Fig.~\ref{fig:Triang}.

The main difference from model $A$ is that we introduce ``indirect''
hopping between adjacent edges which are \emph{angle insensitive}.  We
regard the vertices of $\mathcal{T}$ as additional \emph{virtual}
sites at a higher chemical potential through which a particle can hop.
The benefit is that the type $3$ move becomes symmetric: the
$60^\circ$, $60^\circ$ hops cancel the $120^\circ$, $120^\circ$ hops.
We also must include some ``direct'', $60^\circ$ only, hopping terms
to avoid killing the desired processes 1 and 2.  Here are the
tunnelling amplitudes (again setting $U=1$).

\noindent\underline{indirect:} There will be a tunnelling
amplitude $t$ between any two edges of $T$ which share a vertex
(regardless of color or angle) $t=\epsilon$.

\noindent\underline{direct:}  $t^g_{bb} = 0$, $t^b_{rb} =c_1 \epsilon$,
$t^b_{gb}=c_2 \epsilon$, $ t^r_{bb} = c_3 \epsilon$.

\noindent\underline{Ring term:}
 $\left( \begin{smallmatrix}
   c_4 b \epsilon^2 & -c_4 \epsilon^2  \\
  -c_4 \epsilon^2 & c_4 b^{-1} \epsilon^2
  \end{smallmatrix}\right)$.

Here are the terms in the Hamiltonian $H$ and the corresponding
equations:\\
Type (1):  
\begin{equation} 
  \begin{pmatrix}
    v^b_{gb} & -2 c_1 c_2 \epsilon^2   \\
    -2c_1 c_2 \epsilon^2 & v^b_{rb}
  \end{pmatrix} \sim
  \begin{pmatrix}
    a & -1  \\
    -1 & a^{-1}
  \end{pmatrix}
\end{equation}
Type (1'):  
\begin{equation}
  \begin{pmatrix}
    v^b_{bb} & -2 c_1c_2 \epsilon^2  \\
    -2c_1 c_2 \epsilon^2 & v^b_{rb}
  \end{pmatrix} \sim
  \begin{pmatrix}
    a & -1  \\
    -1 & a^{-1}
  \end{pmatrix}
\end{equation}
Type (2):  
\begin{multline}
  \begin{pmatrix}
    v^r_{bb} & 
    \overset{(\text{\scriptsize{dir dir}})}{-2c^2_2 \epsilon^2} 
    \overset{(\text{\scriptsize{dir ind}})}{- 4c_2 \epsilon^2} 
    \overset{(\text{\scriptsize{ind ind}})}{-0}  \\
    -2 c^2_2 \epsilon^2 -4 c_2 \epsilon^2 & v^r_{bb}
  \end{pmatrix} 
  \\
  \sim
  \begin{pmatrix}
    1& -1  \\
    -1 & 1
  \end{pmatrix}
\end{multline}
Type (3):  
\begin{equation}
  \begin{pmatrix}
    v^g_{bb} & 0 \\
    0 & v^g_{bb}
  \end{pmatrix} =
  \begin{pmatrix}
    0 & 0  \\
    0 & 0
  \end{pmatrix}
\end{equation}
Type (5):
\begin{equation}
  \quad \text{ Ring}  \sim   
  \begin{pmatrix}
    b c_4 \epsilon^2 & -c_4 \epsilon^2   \\
    -c_4 \epsilon^2 & b^{-1}c_4 \epsilon^2
  \end{pmatrix}
\end{equation}
We must therefore impose the following equations:
\begin{equation}
  \begin{array}{ll}
    { } &      {a^4}/{b} ={d} \\
    (2):  & v^r_{bb} = (2c^2_2 + 4 c_2 )\epsilon^2 \\
    (1) \text{ and } (1'):  &  v^b_{gb} = v^b_{bb} = 2ac_1 c_2
    \epsilon^2 \\
    { } & \text{ and } v^b_{rb} = v^b_{rg}= 2a^{-1}
    c_1 c_2 \epsilon^2\\
    (3): & v^g_{bb} = 0
  \end{array}
\end{equation}
As in bosonic models, the required potentials can be solved for from
the tunnelling amplitudes.

As with model $A$, we may simply omit the Ring term from model $C$ to
get model $C^-$.  Under the assumption that an
effective ring term with $b=1$ emerges, $C^-$ could be tuned to yield
$\mathcal{D}K_k$.

Finally we treat model $D$.  We use the same $\mathcal{R}$, red
sublattice \emph{with} defects as in model $B$.  The necessary
colorings are illustrated in Fig.~\ref{fig:model_D}.
\begin{figure}[bht]
\includegraphics[width=2.75in]{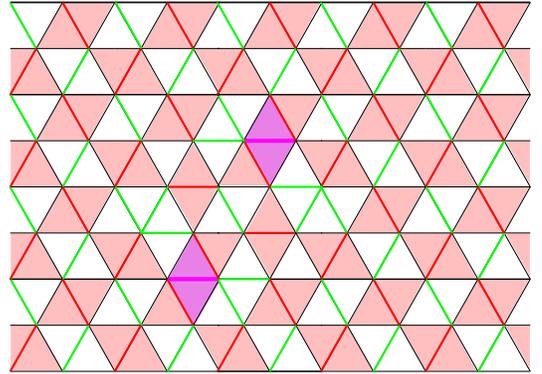}
\caption{Sublattice defect in fermionic model $D$.}
\label{fig:model_D}
\end{figure}

We list the tunnelling amplitudes:

\underline{indirect:} $t=\epsilon$
between all adjacent edges, $e_i \cap e_j \neq 0$.

\underline{direct:}
\begin{subequations}
  \label{eq:direct_tunn_amp}
  \begin{eqnarray}
    & &   0  = t^g_{bb}= t^g_{gb} \\
    & &   c_1 \epsilon = t^b_{rb} =  t^b_{lb} \\
    & &   c_2 \epsilon = t^b_{gb} =  t^b_{gg} \\
    & &   c_3 \epsilon = t^r_{bb} =  t^r_{lb} \\
    & &   c_4 \epsilon = t^l_{rb}
  \end{eqnarray}
\end{subequations}
We have only listed colors $b, r, g, l$ but some of these may appear
in two or three forms because of differing local environment.  Though
each of these forms needs an individual $\mu$ (chemical potential),
the $t$ and $v$ symbols will depend only on the 4 given colors.

Near the defect, to maintain the equations of model $A$ we need to
set: $v^g_{gb} =v^g_{bb}$, $v^r_{bl} =v^r_{bb}$, $v^b_{gl}= v^b_{gb}$,
$v^b_{rl}= v^b_{rb}$, $v^b_{lb} =v^b_{bb}$.

This effectively maintains the constraints associated to rhombi of
types 1, $1^{'}$, 2, and 3 near the defect.

The lavender rhombi (Fig.~\ref{fig:model_D}) yield a new term which
must be constrained by the equations below.
\begin{equation}
    \begin{pmatrix}
        v^l_{bb} & -2c^2_{4} \epsilon^2  \\
       -  2c^2_4 \epsilon^2 & v^l_{rr}
    \end{pmatrix} \sim
    \begin{pmatrix}
         b & {-1}  \\
         {-1} & b^{-1}
    \end{pmatrix}, \quad \frac{a^2}{b} =d
\end{equation}
so we have the equations:
\begin{equation}
  v^l_{bb}= 2b c^2_4 \epsilon^2, \quad v^l_{rr} =2b^{-1} c^2_4
  \epsilon^2, \quad \frac{a^2}{b} =d,
\end{equation}
in addition to the equation from model $C$:
\begin{subequations}
  \label{eq:v-terms}
  \begin{eqnarray}
    & &  v^r_{bl} = v^b_{bb}= (2c^2_2 + 4c_2) \epsilon^2 \\
    & &  v^b_{gl} = v^b_{gb} =  v^b_{bb}= 2a c_1 c_2 \epsilon^2 \\
    & &  v^b_{rl} = v^b_{rb} =  2a^{-1} c_1 c_2 \epsilon^2 \\
    & &  v^g_{gb} = v^g_{bb} = 0.
  \end{eqnarray}
\end{subequations}

To find solutions it is only necessary to choose $a$ and $b$
compatible, pick the positive constants $c_1, \cdots , c_4$ and
compute the potentials $v$.

\begin{acknowledgments}
  The authors would like to thank D.~Jetchev for kindly providing a
  computer program for exploring dimer dynamics and K.~Walker for his
  help in automated diagrammatic calculation of the Jones-Wenzl
  projectors reproduced in section \ref{sec:d-isotopy}. The authors
  would also like to thank S.~Kivelson, S.~Sondhi, K.~Walker, and
  Z.~Wang for helpful discussion and the Aspen Center for Physics
  where a part of this paper was completed for its hospitality. C.N.
  acknowledges the support of the National Science foundation under
  grant DMR-9983544 and the Alfred P. Sloan Foundation.
\end{acknowledgments}

\appendix
\section{Definition and Properties of $\mathcal{D}K_k$} 
\label{appendix1}

The information in this appendix allows one to predict the outcome of
any Aharonov-Bohm experiment conducted on a material thought to be in
the topological phase $\mathcal{D}K_k$.

There is a well known \cite{Witten89} family of topological quantum
field theories (TQFTs) $\text{SU}(2)_k$.  These are the unitary
$\text{SU}(2)$ TQFTs at level $k$.  In the quantum group approach
\cite{Turaev} in order to define precisely this theory (and not e.g.
its time reverse) one sets the deformation parameter $A$ for the
universal enveloping algebra of $su(2)$ to $A= \text{e}^{2 \pi
  \text{i}/4(k+2)}$.

$\text{SU}(2)_k$ has has particles species indexed by $\text{SU}(2)$
representations (or `isospins') $j=0,1/2,1,\ldots,k/2$.  These $k+1$
irreducible representations are subject to the following rules:

The fusion space $V^{{j_3}}_{{j_1}{j_2}} \cong \mathcal{C}$ if
\begin{subequations}
  \label{eq:fusion_space}
  \begin{eqnarray}
    & & 0 \leq {j_1}, {j_2}, {j_3}, \leq k/2 \\
    & & {j_1} +{j_2} \geq {j_3},\\
    & & {j_3} + {j_1} \geq {j_2}, \\
    & & {j_2} + {j_3} \geq {j_1}, \\
    & & {j_1} + {j_2} + {j_3} \leq k.
  \end{eqnarray}
\end{subequations}
$\cong 0$ otherwise.

The $S-$matrix is
\begin{multline}
  \label{eq:A2}
  S_{{j_1}{j_2}} =\sqrt{\frac{2}{k+2}}
  \big(\sin \,(2{j_1}+1)(2{j_2}+1) \,\pi/(k+2)\big),\\
  0<i, j \leq k/2.
\end{multline}

There is a variant $K_k$ of $\text{SU}(2)_k$ worked out in
\cite{Blanchet95} (also see Refs.~\onlinecite{Good_guys,FNWW}) based
on the Kauffman bracket for links but now with $A=\text{i}\text{e}^{2
  \pi \text{i} /4(k+2)}$ in the relation of Fig.~\ref{fig:Kauffman}.
\begin{figure}[hbt]
\includegraphics[width=1.5in]{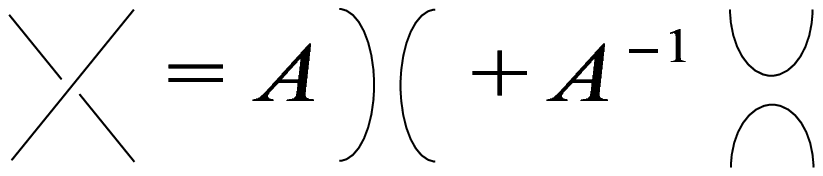}
\caption{Kauffman bracket}
\label{fig:Kauffman}
\end{figure}

On a torus $T^2$, $K_k (T^2 )$ would be a subspace of the functions on
links in the \emph{solid} torus, respecting the above relation, the
level $k$ Jones-Wenzl relation, small unknots $= d= 2 \cos \pi/k+2$,
and finally the relation that a right handed kink is equal $-A^{-3}$.
$K_k$ also has $k+1$ orthogonal irreps $0, \ldots , k$ with the same
fusion relations as $\text{SU}(2)_k$ but the $S-$ matrix is
$\widetilde{S}_{{j_1}{j_2}} = (-1)^{4{j_1}{j_2}} \sqrt{\frac{2}{k+2}}
\big( \sin (2{j_1}+1) (2{j_2}+1) \pi/(k+2) \big)$.

$K_k$ is a Chern-Simons-Higgs theory in which two $\text{SU}(2)$
Chern-Simons gauge fields, with levels $k$ and $1$, are tied together
by a condensate which transforms in the isospin $k/2$ representation
of the first and the isospin $1/2$ representation of the second:
\begin{equation}
{\cal S} = k {\cal S}_{\text{CS}}\left[{a_1}\right] - 
{\cal S}_\text{CS}\left[{a_2}\right] 
+ {\cal S}_\text{Higgs}[\Phi,{a_1},{a_2}]
\end{equation}
where
\begin{eqnarray}
{\cal S}_\text{CS} &=& \frac{1}{4\pi} \int
{\epsilon^{\mu\nu\rho}}\left({a_\mu^{\underline a}}
{\partial_\nu}{a_\rho^{\underline a}}
+ \frac{2}{3}\,{f_{{\underline a}\,{\underline b}\,{\underline c}}}
{a_\mu^{\underline a}}{a_\nu^{\underline b}}{a_\rho^{\underline c}}\right)
\cr
&=& \frac{1}{4\pi}\int \text{tr}\left(a\wedge da
+ \frac{2}{3} a\wedge a\wedge a\right)
\end{eqnarray}
and
\begin{multline}
  {\cal S}_\text{Higgs}[\Phi,{a_1},{a_2}] = \int
  \left|\left(\text{i}{\partial_\mu} + a_{1\mu}^{\underline a} 
      T^{\underline a}_{IJ} + 
      a_{2\mu}^{\underline a} t^{\underline a}_{ij} \right)
    {\Phi_{Jj}} \right|^2
  + V(\Phi)
\end{multline}
${\underline a}=1,2,3$ are $su(2)$ Lie algebra indices; $T^{\underline
  a}$, $t^{\underline a}$ are $su(2)$ generators in the isospin $k/2$
and $1/2$ representations, respectively.  $V(\Phi)$ is minimized at
non-zero $|\Phi |$. The presence of the condensate restricts the
possible particle types so that half-integer isospins under
$\text{SU}(2)_k$ are necessarily spin-$1/2$ under $\text{SU}(2)_1$.
This leads to the additional minus signs in the braiding statistics
and the $S-$matrix.

For $k$ even, det $\widetilde{S} =1$ but for $k$ odd det
$\widetilde{S} = 0$. This singularity is explained by the isomorphism:
\begin{equation}
  \label{eq:A3}
  \widetilde{S}_{{j_1}{j_2}} \cong
  \left( \begin{smallmatrix}
      1 & 1 \\
      1 & 1
    \end{smallmatrix}\right)  \otimes \widetilde{S}_\text{integer, integer} 
  \text{ for } k \text{ odd }.
\end{equation}

Thus $K_k$ is only a true TQFT for $k$ even.  Physically, the problem
with odd $k $ is that it is not possible to distinguish the
isospin-$j$ particle from the isospin-$(k/2 - j)$ particle by any
braiding operation \cite{Good_guys,FNWW}.

Planar arc diagrams lead both to $K_k$, via Fig.~\ref{fig:Kauffman},
and to $\text{SU}(2)_k$, via the Rumer-Teller-Weyl theorem on
$\text{SU}(2)$ representations (see Ref.~\onlinecite{Kuperberg96} for
an exposition of both). The subtle difference is that certain $-1$'s,
Frobenius-Schur indicators, occur in the representation theory coming
from the fact that the fundamental representation of $\text{SU}(2)$ is
\emph{antisymmetrically} self-dual, whereas these signs are absent in
Kauffman's theory.

We come now to $\mathcal{D}K_k$, the Drinfeld double of $K_k$.  It is
a subspace of functions on links in surface $\times$ interval (same
relations as listed for solid torus) and has its own naturally defined
positive definite hermitian inner product and a unitary $S-$matrix,
$\widetilde{\widetilde{S}}$ \cite{Blanchet95,BK:book}. When $k$ is
even
\begin{equation}
  \label{eq:A4}
  \mathcal{D}K_k \cong K^{\ast}_{k} \otimes_\mathbb{C} K_k 
  \text{ as Hilbert spaces, and }
  \widetilde{\widetilde{S}} = \widetilde{S}\otimes
  \widetilde{S},
\end{equation}
when $k$ is odd, we still have:
\begin{multline}
  \label{eq:A5}
  \mathcal{D}K_k \cong K^{\ast}_{k} 
  \otimes_{\mathbb{C}} K_k,
  \\
  \text{ but now }
  \widetilde{\widetilde{S}} = T \otimes 
  \widetilde{S}_\text{integer, integer}
  \otimes \widetilde{S}_\text{integer, integer},
\end{multline}
i.e. $\widetilde{\widetilde{S}}$ is the tensor square of the
nonsingular even part of $\widetilde{S}$ tensor $T$, the $4 \times 4
\,\,S$ matrix of the level $k=1$ theory $\mathcal{D}K_1$, In Kitaev's
``toric code" notation \cite{Kitaev97} the irreps of $\mathcal{D}K_1$
are: $\emptyset, e, m, $ and $em$; the trivial, electric, magnetic,
and electric and magnetic excitations respectively. The Lagrangian for
this sector is that of $\text{Z}_2$ theory \cite{Wegner71}.

In this basis:
\begin{equation}
  \label{eq:A6}
  T =
  \begin{array}{cc}
    { } & 
    \begin{matrix}
     \;\; \emptyset & \quad\;\;  e & \quad\;\;  m & \quad\;\;  em
    \end{matrix} 
    \\
    \begin{matrix}
      \emptyset \\
      e \\
      m \\
      em \,
    \end{matrix}
    &
    \begin{pmatrix}
      1/2 & 1/2 &  1/2 &  1/2 \\
      1/2 & 1/2 &  -1/2 & - 1/2 \\
      1/2 & -1/2 & 1/2 &  -1/2 \\
      1/2 & -1/2 & -1/2 & 1/2
    \end{pmatrix}
  \end{array}
\end{equation}

The isomorphisms in Eqs.~(\ref{eq:A3})--(\ref{eq:A5}) can be made
explicit as follows (see Refs.~\onlinecite{Good_guys,FNWW}).

$K_k$ has basis $0, \ldots, k/2$.  Define $\widehat{i} = k-i$,
$0\leq i \leq K$.  If $k$ is odd then exactly one of $i$ and
$\widehat{i}$ is even; denote that one by $i_e$.  So when $k$ is
odd we write $i= (i_e , \sigma_i)$ where $\sigma_i = +$ or $-$ as
$i$ is even or odd so $(i_e, +) = i_e$, $(i_e , - ) =
\widehat{i}_e$.

$\mathcal{D}K_{k}^{\ast} \otimes \mathcal{D}K_k$ has basis $(i, j, l,
m)$, $0 \leq i, j, l, m \leq k/2$, and \break $(i, j, l, m) = ( i_e ,
\sigma_i , j_e , \sigma_j , l_e , \sigma_l , m_e, \sigma_m)$ in the
case $k=$ odd. Declare an isomorphism $\theta$ from $(\sigma_i,
\sigma_{j})$ and from $(\sigma_l , \sigma_m )$ to the toric code
excitations by: $(+, +) \to \emptyset , (-, +) \to e, (+, -) \to m, $
and $(-, -) \to em$. Then to realize (\ref{eq:A5}), map $(i,j, l, m)$
to $\big( \theta (\sigma_i , \sigma_j), \theta (\sigma_l , \sigma_m),
i_e , j_e , l_e, m_e \big)$.

\section{Nearly degenerate perturbation theory}
\label{appendix2}

Although for the purposes of this paper we do not use anything beyond
the second-order perturbation theory, here we present a general scheme
useful for extending this type of analysis to higher orders. In
particular, three and four-dimer moves will appear as higher order
terms in the effective Hamiltonian.

The perturbative scheme presented here closely follows that
developed in Ref.~\onlinecite{MacDonald88}. The idea is to recursively
block-diagonalise the Hamiltonian, order by order eliminating terms
that change the number of dimer collisions (a collision occurs when
two dimers share a vertex). The resulting effective Hamiltonian then
contains terms (up to a given order) that only connect states within
sectors with a fixed number of such collisions. To proceed with this
programme, we rewrite the original Hamiltonian (\ref{eq:Hubbard}) as
\begin{equation}
  \label{eq:Ham_split}
  H = T_0 + T_{1} + T_{-1} + T_{X} + T_{-X} + D + W .
\end{equation}
Here $ T_0 + T_{1} + T_{-1} + T_{X} + T_{-X}= -\sum_{\langle
  i,j\rangle} t_{ij}\left(c^\dag_i c_j + \text{h.c.}\right)$, $D=
\sum_{i} \mu_i n_i + \sum_{(i,j)\in \bowtie, \notin \hexagon } V_{ij}
n_i n_j$ is the ``low-energy'' part of potential energy, while $W$
combines the remaining ``high-energy'' terms in $H$, i.e.  all dimer
collision interactions.  $T_0$, $T_{1}$ and $T_{-1}$ represent the
dimer moves that respectively do not change, increase by one, or
decrease by one the number of collisions. $T_{X}$ results in a dimer
ending on top of another dimer (double occupancy in the original
particle language), while $T_{-X}$ undoes that. Since we have already
chosen $U_0=\infty$ thus restricting our attention to the
single-occupancy subspace within which $T_{\pm X}$ have zero matrix
elements, we should drop these terms from our consideration.  Notice
that $T_{m}^\dag=T_{-m}$ and $[T_m,W]= - m U T_m$, $m=0, \pm 1$.

Our strategy is to recursively construct the operator $\text{i} S$
such that a ``rotated'' Hamiltonian $\tilde{H}$ given by
\begin{equation}
  \label{eq:unitary}
  \tilde{H} = \text{e}^{\text{i} S} H \text{e}^{-\text{i} S} = H +
  [\text{i} S,H] + \frac{1}{2!}[\text{i} S,[\text{i} S,H]] + \ldots
\end{equation}
is block-diagonal as described above. As a first approximation, we
choose $\text{i} S^{(1)} = (T_1-T_{-1})/U$ which leads to $H^{(2)}=
T_0+D+W +\left([T_1,D]-[T_{-1},D] +
  [T_1,T_0]-[T_{-1},T_0]-[T_{-1},T_1]\right)/U +
\mathcal{O}\left(x^3/U^2\right)$, where
$x=\max\left\{t_{ij},V_{ij},\mu_i\right\}$ (here we follow the
numbering convention of Ref.~\onlinecite{MacDonald88} where
$H^{(1)}\equiv H$).  This procedure eliminates $T_1$ and $T_{-1}$ but
generates a slew of smaller terms, all of which, except
$-[T_{-1},T_1]/U$, change the number of dimer collisions.  The next
step is to correct $\text{i} S$ in order to eliminate these new
unwanted terms, repeating this procedure recursively up to any given
order. The number of terms rapidly gets out of hand, and we used
Mathematica to keep track of them up to the fourth order (i.e. keeping
terms of order $x^4/U^3$).

We are particularly interested in the terms that connect states within
the lowest energy, zero collisions sector. Thus we can write
$\tilde{H}=\tilde{H}'+ \tilde{H}''$ where $\tilde{H}''$ consists of
terms vanishing in this sector (i.e. the matrix element $\langle
\alpha | \tilde{H}'' | \beta \rangle = 0$ for any two proper dimer
coverings $|\alpha\rangle$ and $|\beta\rangle$). Then the effective
low-energy Hamiltonian is given to the fourth order by:
\begin{multline}
  \label{eff_Ham_1}
  \tilde{H}^{\prime (4)}
  =  D - {T_{-1}}{T_{1}}/U
  \\
  - \left(\frac{1}{2}\, D {T_{-1}}{T_{1}} + {T_{-1}} D {T_{1}}
  - \frac{1}{2}\, {T_{-1}}{T_{1}} D
  + {T_{-1}}{T_{0}}{T_{1}}\right)/U^2
  \\
  + \bigg(-\frac{1}{2}\, D D  {T_{-1}} {T_{1}} + D {T_{-1}}D {T_{1}}
  - {T_{-1}}D D {T_{1}}
  \\
  +  {T_{-1}}D {T_{1}}D
  - \frac{1}{2}\, {T_{-1}} {T_{1}} D D + D {T_{-1}} {T_{0}} {T_{1}}
  \\
  - {T_{-1}}D {T_{0}} {T_{1}} - {T_{-1}} {T_{0}}D {T_{1}}
  + {T_{-1}} {T_{0}} {T_{1}}D
  \\
  - {T_{-1}} {T_{0}} {T_{0}} {T_{1}}
  + {T_{-1}} {T_{1}} {T_{-1}} {T_{1}}
  - \frac{1}{2}\, {T_{-1}} {T_{-1}} {T_{1}} {T_{1}}\bigg)/U^3
  \\
  = D -  \frac{{T_{-1}}{T_{1}}}{U}
  + \frac{1}{2U^2} \big\{{T_{-1}} \left[D, {T_{1}}\right] +
    \text{h.c.}\big\}
  + \frac{{T_{-1}}{T_{0}}{T_{1}}}{U^2}
  \\
  - \bigg(\frac{1}{2}
  \big\{{T_{-1}}\left[D, \left[D, {T_{1}}\right]\right]
    + \text{h.c.}\big\}
  -\big\{{T_{-1}} {T_{0}} \left[D,{T_{1}}\right] +
    \text{h.c.}\big\}
  \\
  - {T_{-1}} {T_{0}} {T_{0}} {T_{1}}
  + {T_{-1}} {T_{1}} {T_{-1}} {T_{1}}
  - \frac{1}{2}\, {T_{-1}} {T_{-1}} {T_{1}} {T_{1}}\bigg)/U^3
\end{multline}

The advantage of rewriting this Hamiltonian using commutators such as
$[D,T_1]$ becomes clear if we recall that $D=\sum_{i} \mu_i n_i +
\sum_{(i,j)\in \bowtie, \notin \hexagon } V_{i j} n_i n_j$ is diagonal
in the local position basis. This allows us to combine the terms as
follows:
\begin{multline}
  \langle \alpha\left| \left(
      - \, {T_{-1}}{T_{1}}
      + \frac{1}{2} \left({T_{-1}} \left[D, {T_{1}}\right]
        + \text{h.c.}\right)\right. \right.
  \\
  \left.\left. - \frac{1}{2}
      \left({T_{-1}}\left[D, \left[D, {T_{1}}\right]\right]
        + \text{h.c.}\right)\right)\right|\beta \rangle
  \\
  = - \frac{1}{2} \bigg\{\sum_n
  \left(1-\frac{d_n-d_\beta}{U}
    +\left(\frac{d_n-d_\beta}{U}\right)^2\right)
  \\
  \times \langle\alpha\left|{T_{-1}}\right| n\rangle
  \langle n\left|{T_{1}}\right| \beta\rangle
  + (\alpha \leftrightarrow \beta) \bigg\}
  \label{eq:term_comb1}
\end{multline}
and
\begin{multline}
  \langle \alpha\left| \left(
      {T_{-1}}{T_{0}}{T_{1}}
      -\left({T_{-1}} {T_{0}} \left[D,{T_{1}}\right] +
        \text{h.c.}\right) \right)\right|\beta \rangle
  \\
  =  \sum_n \left(\frac{1}{2}-\frac{d_n-d_\beta}{U}\right)
  \langle\alpha\left|{T_{-1}}{T_{0}}\right| n\rangle
  \langle n\left|{T_{1}}\right| \beta\rangle
  \\
  + (\alpha \leftrightarrow \beta)
  \label{eq:term_comb2}
\end{multline}
where $|n\rangle$ is an excited state with one dimer collision and
$d_n$, $d_\beta$ are the eigenvalues of $D$. Written in form
(\ref{eq:term_comb1}),(\ref{eq:term_comb2}), these terms explicitly
depend only on local differences in the two dimer configurations
rather than the expectation values of infinite sums $D$.


\end{document}